\documentclass[twocolumn]{aastex631}

\usepackage{times}
\usepackage{orcidlink}

\def\d3{$\delta_{3}$ }
\def\1d3{$(1 + \delta_{3})$ }
\def\l1d3{$\log_{10}(1 + \delta_{3})$ }

\def\s3{$\Sigma_{3}$}
\def\hamath{\mathrm{H}\alpha}
\def\ha{H$\alpha$}
\def\hb{H$\beta$}

\def\ntwo{[NII]}
\def\stwo{[SII]}

\def\24m{24 $\mu$m}

\def\sm{$\rm~M_{*}$}

\def\kms{${\rm km~s^{-1}}$ }

\def\Msolar{$\rm M_{\odot}$}

\def\rmxaa{RMxAA}

\def\sigsm{$\Sigma_{*}$}
\def\sigsfr{$\Sigma_{\rm SFR}$}

\def\h2{$\rm H_{2}$}

\def\Mh2{$\rm M_{H_{2}}$}

\def\sigh2{$\Sigma_{\rm H_{2}}$}

\def\fgas{$f_{\rm gas}$}
\def\fh2{$f_{\rm H_{2}}$}

\def\Re{$R_{e}$}
\def\co{$^{12}$CO(1-0)}

\def\taudecay{$\tau_{decay}$}

\shorttitle{Multiple paths in quenching green valley galaxies}
\shortauthors{Lin et al.}

\begin{document}

\title{The ALMaQUEST Survey XVII: Unveiling Multiple Quenching Pathways in Green Valley Galaxies via Molecular Gas and Quenching Timescale Analyses}

\author{Lihwai Lin}
\altaffiliation{Email: lihwailin@asiaa.sinica.edu.tw}
\affiliation{Institute of Astronomy \& Astrophysics, Academia Sinica, No.1, Sec. 4 Roosevelt Road, Taipei 10617, Taiwan}


\author{Po-Feng Wu}
\affiliation{Graduate Institute of Astrophysics and Department of Physics, National Taiwan University, No.1, Sec. 4 Roosevelt Road, Taipei 10617, Taiwan}

\author{Mallory D. Thorp}
\affiliation{Argelander-Institut für Astronomie, Universität Bonn, Auf dem Hügel 71, 53121 Bonn, Germany}

\author{Asa F. L. Bluck \orcidlink{0000-0001-6395-4504}}
\affiliation{Stocker AstroScience Center, Dept. of Physics, Florida International University, 11200 S.W. 8th Street, Miami, FL 33199, USA}

\author[0000-0002-1370-6964]{Hsi-An Pan}
\affiliation{Department of Physics, Tamkang University, No.151, Yingzhuan Road, Tamsui District, New Taipei City 251301, Taiwan}

\author{Sara L. Ellison}
\affiliation{Department of Physics \& Astronomy, University of Victoria, Finnerty Road, Victoria, British Columbia, V8P 1A1, Canada}

\author{Kate Rowlands}
\affiliation{Space Telescope Science Institute, 3700 San Martin Dr Baltimore, MD 21218, USA}

\author{Justin Atsushi Otter}
\affiliation{William H. Miller III Department of Physics and Astronomy, Johns Hopkins University, Baltimore, MD 21218, USA}

\author{Sebasti\'{a}n F. S\'{a}nchez }
\affiliation{Instituto de Astronom\'ia, Universidad Nacional Aut\'onoma de  M\'exico, Circuito Exterior, Ciudad Universitaria, Ciudad de M\'exico 04510, Mexico}

\begin{abstract}
	
Statistically, green valley (GV) galaxies exhibit lower molecular gas fractions ($f_{\mathrm{gas}}$) and reduced star formation efficiency (SFE) compared to star-forming galaxies. However, it remains unclear whether quenching is primarily driven by one factor or results from a combination of mechanisms in individual GV galaxies. In this study, we address this question by examining the spatial distributions of star formation and molecular gas in 28 GVs selected from the ALMaQUEST survey and additional literature samples. For each galaxy, we identify regions with suppressed specific star formation rate (sSFR) and measure $\Delta f_{\mathrm{gas}}$ and $\Delta$SFE—offsets from the resolved scaling relations of the star-forming main sequence galaxies. By comparing the fraction of regions with negative $\Delta f_{\mathrm{gas}}$ and $\Delta$SFE, we classify 35.7$\pm$13.2\% (57.1$\pm$17.9\%) of GV galaxies as $f_{\mathrm{gas}}$-driven, 39.3$\pm$14.0\% (39.3$\pm$14.0\%) as SFE-driven, and 25.0$\pm$10.6\% (3.6$\pm$3.6\%) as mixed mode when adopting a fixed (variable) CO-to-\h2 conversion factor ($\alpha_{\mathrm{CO}}$). These results indicate that GVs undergo quenching through multiple pathways. As sSFR decreases from the main sequence to the green valley, we observe a transition toward predominantly SFE-driven quenching, possibly linked to internal processes such as morphological quenching or AGN activity. We further estimate the quenching timescale (\taudecay), defined as the time from the peak SFR to 1/e (approximately 37\%) of its value, using integrated MaNGA spectra. SFE-driven quenching is typically associated with short \taudecay, while $f_{\mathrm{gas}}$-driven quenching shows a broader range. Overall, 75\% of GVs exhibit \taudecay~ shorter than 1 Gyr, suggesting that quenching in most GVs proceeds rapidly, challenging purely slow-quenching scenarios like starvation.

\end{abstract}


\section{INTRODUCTION}

Green Valley (GV) galaxies represent an intermediate population located between the blue cloud and red sequence \citep[e.g.,][]{mar07,sal07,wyd07,sal14,coe18}. Although historically GV galaxies were identified based on their restframe optical colors,  a popular way of selecting GV galaxies nowadays is to identify the population lying between the star-forming main sequence (SFMS) and quiescent galaxies on the global star formation rate (SFR) and global stellar mass (\sm) plane \citep[e.g.,][]{bri04,dad07,sai16,lin17,bel18,jia20,bro20,pio20,lin22,jia22,wan22}. These galaxies are likely in the transition from the star-forming phase to the quiescent state, and hence offer valuable insights into the underlying physical mechanisms driving quenching.

Numerous studies have investigated the abundance and physical conditions of molecular gas to understand how and why star formation is suppressed in GV galaxies \citep{lin17,bro20,lin22,vil24,lin24}. For instance, global measurements of gas properties extending to GV and even fully quenched galaxies show systematically lower star formation efficiency (SFE $\equiv$ SFR/\Mh2) and molecular gas fraction (\fgas $\equiv$ \Mh2/\sm) in these populations \citep{lin20,vil24,col25}, where \Mh2~denotes the molecular gas mass. Using the spatially resolved gas properties of nearby GV galaxies from the ALMA-MaNGA QUEnching and STar formation (ALMaQUEST) survey \citep{lin20,ell24}, \citet{lin22} found that these galaxies systematically deviate from main sequence (MS) galaxies across three key scaling relations: the resolved star-forming main sequence (rSFMS), the resolved Schmidt–Kennicutt relation (rSK), and the resolved molecular gas main sequence (rMGMS) \citep[e.g.,][]{lin19b,mor20,san21,ell21b,pes21,bak22}. \citet{lin22} further reported that both the resolved  SFE and the resolved \fgas~ are statistically lower in GV galaxies compared to their MS counterparts \footnote{SFE is defined using the surface densities as \sigsfr/\sigh2~ and \fgas~ is defined as \sigh2/\sigsm~when referring to resolved measurements.}. On the simulation side, the impact of gas depends on the feedback prescriptions. While Illustris quenches star formation primarily by depleting cold gas, Illustris-TNG predicts quenching through a reduction in SFE \citep[e.g.,][]{pio22}.

However, most of the observational measurements were based on either integrated gas properties or ensembles of spaxels, which do not reveal whether the suppression of SFE and \fgas~ occurs uniformly across individual galaxies or varies significantly from one system to another. For example, some galaxies may undergo quenching driven primarily by reduced SFE, while others may be influenced more by low \fgas, or by a combination of both. A recent step forward was made by \citet{pan24}, who analyzed the radial profiles of specific star formation rate (sSFR), SFE, and  \fgas~in individual ALMaQUEST galaxies. Their results suggest that the dominant quenching mechanism—whether SFE-driven or \fgas-driven—can vary with galactocentric radius.

In addition to the gas properties, the quenching mechanism for GV galaxies can also be inferred from the quenching timescale. It has been suggested that the quenching timescales of green valley galaxies vary with morphology and environment \citep{sch14,sme15,jia20,jia22}, reflecting the diversity of various quenching mechanisms. Despite the fact that the quenching timescale has been intensively studied in both simulations and observations, how it links to the gas properties remains unexplored. Quenching processes that involve the removal of gas and hence lower \fgas~through violent physical mechanisms, such as ram pressure stripping and quasar-mode AGN feedback, typically occur within 1 Gyr \citep[e.g.,][]{di05,kav11,muz14,her19}. Starvation \citep{lar80, pen15}, an environmental process that prevents the hot gas from cooling, can operate over a timescale longer than 2 Gyr \citep{bax25}. The timescale of morphological quenching \citep{mar09}, which suppresses SFE by stabilizing molecular gas, varies depending on the mechanisms in transforming the morphology \citep[e.g.,][]{kel19,oxl24}. The AGN radio feedback can both lower \fgas~and suppress SFE by inducing turbulence, operating over a longer timescale \citep{cro06,wei17,ven21}. 
Combining the information of gas properties and the quenching timescale can therefore shed light on quenching mechanisms.

In this work, we adopt the method developed by \citet{tho22} to quantify the relative contributions of SFE and \fgas~for a given galaxy.  For each galaxy, we quantify the fractions of quenched areas attributed to reduced SFE and \fgas. The dominant quenching mode is determined if more than 60\% of the suppressed regions are associated with a particular mode. In addition, we model the star formation histories for the whole sample by performing stellar population analysis using the integrated spectra from MaNGA. We then investigate the distribution of quenching timescales in various quenching modes, which helps discern the potential quenching mechanisms.

In \S2, we describe the sample, the methods of measuring physical quantities, and the spectral fitting procedures used in this work. In \S3, we present the main results on the quenching modes and their quenching timescales. We further discuss the interpretation of our results and limitations of our analyses in \S4. Conclusions are given in \S5. Throughout this paper, we adopt the following cosmology: \textit{H}$_0$ = 70~\kms Mpc$^{-1}$, $\Omega_{\rm m} = 0.3$ and $\Omega_{\Lambda } = 0.7$. We use a Salpeter initial mass function (IMF).

\section{Data, Samples, and Methods}
\subsection{Sample selections}

For the work presented in this study, we include several ALMA programs that observe galaxies taken from the Mapping Nearby Galaxies at Apache Point Observatory \citep[MaNGA;][]{bun15} survey. These include 46 ALMaQUEST galaxies \citep{lin20,ell24} and additional 25 galaxies targeting specific MaNGA-identified subsets, including mergers \citep{tho22} and post-starburst (PSB) galaxies \citep{ott22}.

The ALMaQUEST main sample obtains the \co~distributions for 46 MaNGA galaxies more massive than $10^{10}$\Msolar~ at $0.01<z<0.14$. The survey spans a broad range of sSFR and encompasses starburst (SB), MS, and GV galaxies. The ALMA observations were carried out from ALMA programs 2015.1.01225.S, 2017.1.01093.S, 2018.1.00558.S (PI: Lin), and 2018.1.00541.S (PI: Ellison).  These observations were conducted with a spatial resolution aligned with MaNGA's point spread function (approximately 2.5\arcsec), allowing for a unified analysis of stellar and gas properties at comparable physical scales, ranging from 0.9 to 6 kpc. More details about the sample selection and data reduction are described in the ALMaQUEST survey paper \citep{lin20}. 

To investigate the impact of galaxy-galaxy interaction on the molecular gas properties and its role in regulating star formation activities, 14 close pairs and 6 post-merger galaxies were observed as part of an ALMA Cycle 7 program (2019.1.00260.S, P.I.: Hsi-An Pan), which is referred to as ALMaQUEST-Mergers in this work \citep{tho22}. This set of observations adopts the same methodology used in the ALMaQUEST survey, i.e., \co~mapping with spatial resolution matched to that of the MaNGA survey. For the analyses presented here, we exclude 4 galaxies that are CO undetected. The combined ALMaQUEST and ALMaQUEST-Mergers samples are referred to as the extended ALMaQUEST sample \citep{ell24}.  

Additionally, we include data from the ALMA Cycle 7 program 2019.1.01136.S (P.I. Rowlands), which observed \co~in 14 MaNGA-selected post-starburst (PSB) galaxies. These PSB galaxies were identified using two different methods. The first method selects PSB spaxels based on combined criteria involving H$\delta_A$ absorption and \ha~emission. Specifically, \citet{che19} define PSB regions with H$\delta_A > 3$\AA, W(\ha) $< 10$\AA, and logW(\ha) $<$ 0.23 $\times$ H$\delta_A - 0.46$, where H$\delta_A$ measures the strength of H$\delta$ absorption and W(\ha) represents the equivalent width of \ha. The second method identifies PSB galaxies using the principle component analysis (PCA) approach of \citet{row18}, requiring that more than 50\% of spaxels within the central 0.5Re (where Re is the effective radius) are dominated by PSB spaxels. As such, the second method selects central PSBs while the first method includes both central and ring-like PSBs \citep{che19}. We refer the readers to \citet{ott22} for details on the characteristics of the PSB sample. We exclude 2 undetected sources as well as 3 PSB galaxies that overlap with the extended ALMaQUEST sample. 

\begin{figure}[tbh]
\includegraphics[angle=0,width=0.45\textwidth]{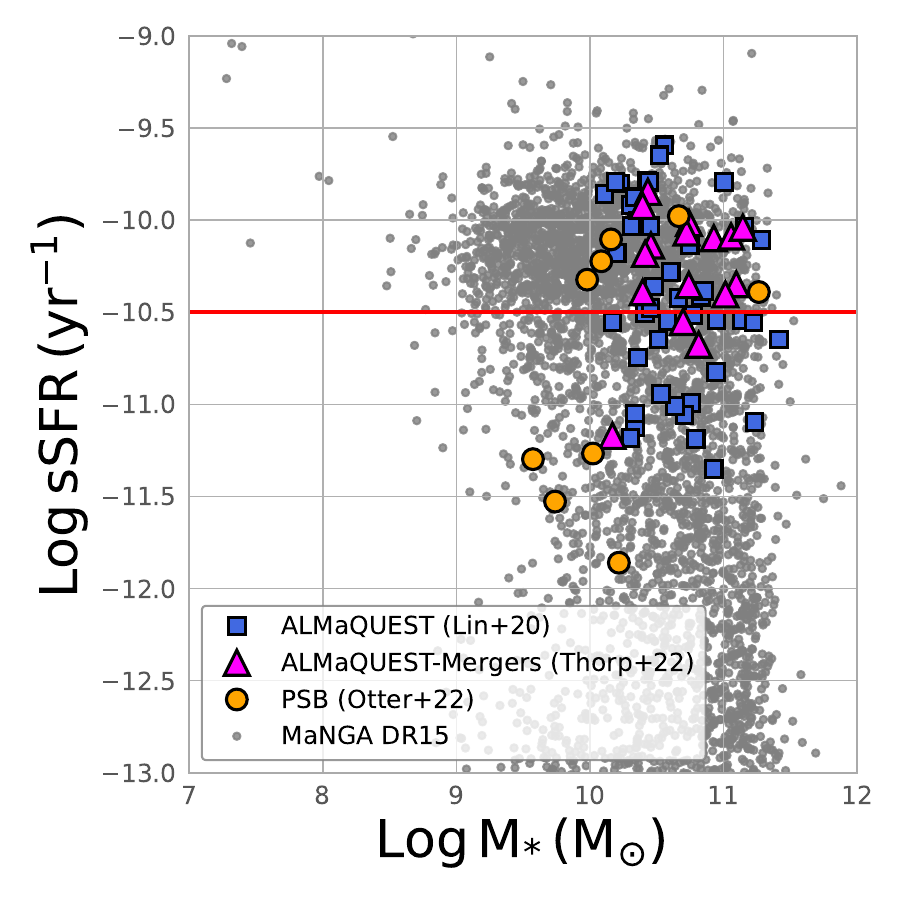}
\caption{The global specific star formation rate (sSFR) vs. stellar mass (\sm) of 71 galaxies used in this work and 4656 MaNGA galaxies (black dots) from SDSS DR15. The global measurements are taken from the Pipe3D \citep{san16a,san16b} value-added catalog \citep{san18} in the SDSS DR15 release \citep{aqu19}. Colored symbols represent various sub-populations, including the ALMaQUEST survey \citep[blue squares;][]{lin20}, ALMaQUEST-Mergers \citep[magenta triangles;][]{tho22}, and post-starburst galaxies \citep[orange circles;][]{ott22}. The red lines denote the dividing line defining the MS and GV subsamples used in this work.  \label{fig:ssf-sm}}
\end{figure}

The final sample used in this study consists of 71 distinct galaxies, which are shown in Figure \ref{fig:ssf-sm}. The basic properties of this sample are displayed in Table \ref{tab:sample}. In this study, we define GV galaxies as those with a global sSFR below $10^{-10.5}$yr$^{-1}$ \citep{lin22}. This leads to 28 galaxies below the sSFR threshold as denoted by the red line in Figure \ref{fig:ssf-sm}. Although the inclusion of galaxy pairs may complicate efforts to disentangle intrinsic and environmental quenching mechanisms \citep[e.g.,][]{vil22}, we retain this sample to remain inclusive of galaxies undergoing a variety of physical processes. As shown in Figure~\ref{fig:ssf-sm}, only three galaxies lie in the green valley, indicating that our sample is not biased toward interacting systems. Likewise, no preselection based on morphology or the presence of an AGN was applied in constructing the GV sample.

\begin{figure*}
\centering
\includegraphics[width=0.95\textwidth]{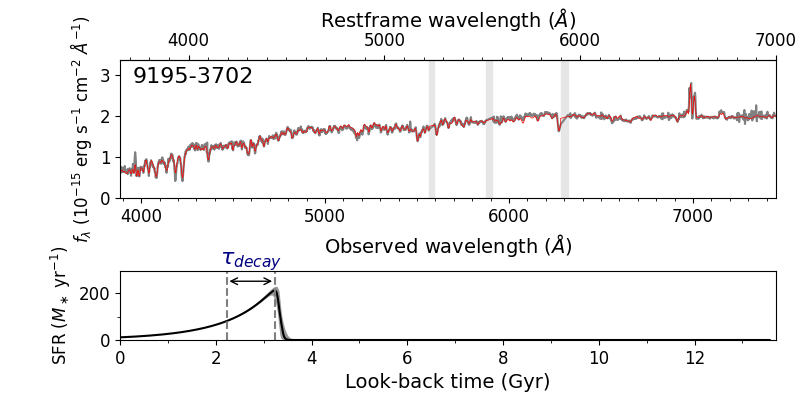}
\caption{An example (MaNGA object with 'plateifu' ID = 9195-3702) of the integrated spectrum (top panel, grey), the model fitting (top panel, red) and the SFH (bottom panel).  The vertical light grey stripes in the top panel indicate the positions of masked sky lines. The black line and the grey shaded areas in the bottom panel show the 50th, 16th, and 84th percentiles of the posteriors. In the bottom panel, the two dashed vertical lines indicate the look-back times corresponding to the peak SFR (right) and the point at which the SFR declines to 1/e of its peak value (left). \label{fig:sfh}}
	
\end{figure*}


\begin{figure*}
\includegraphics[angle=0,width=0.95\textwidth]{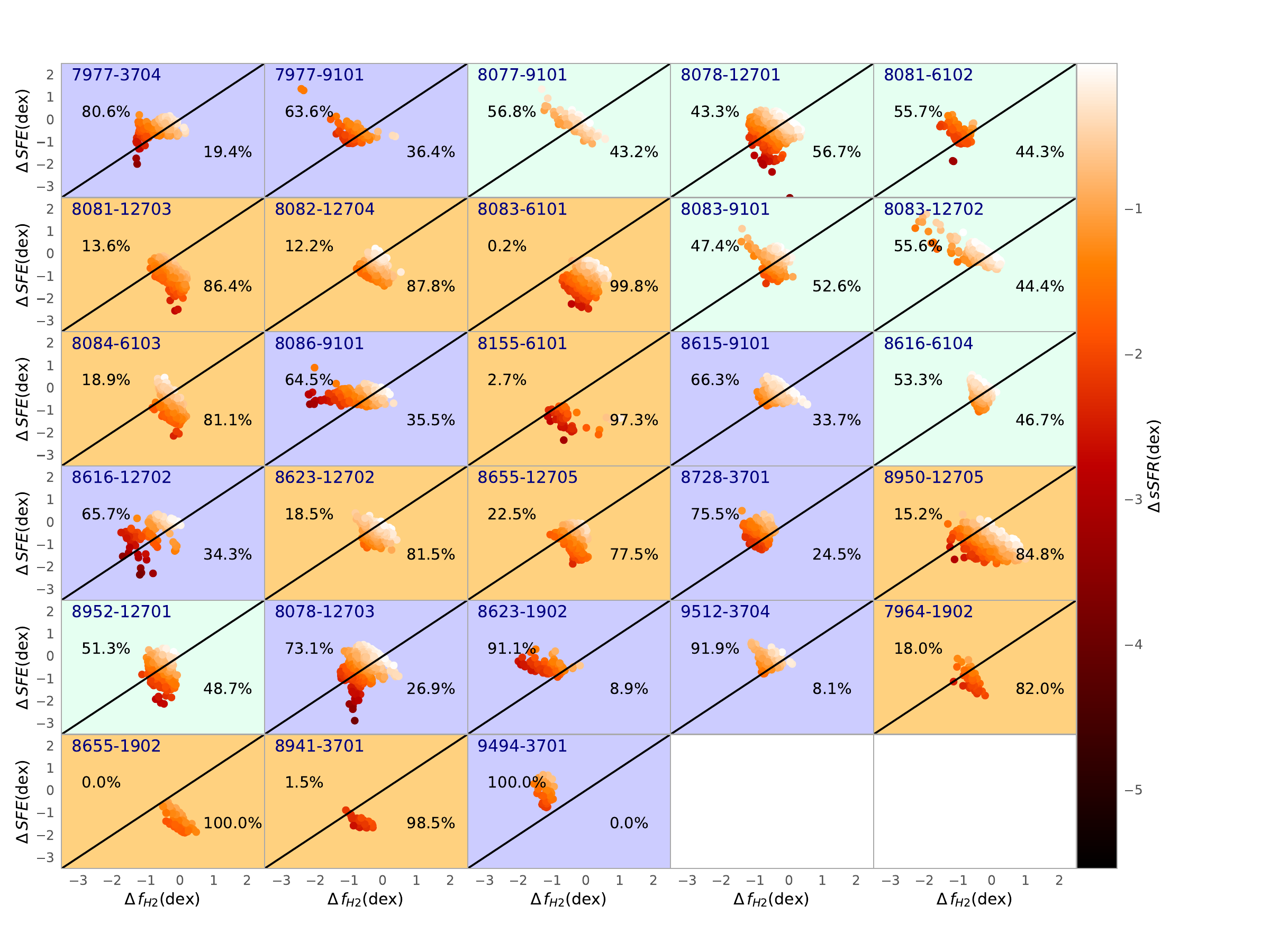}
\caption{$\Delta$SFE versus $\Delta$\fgas~for spaxels with $\Delta$sSFR $<$ 0 for 28 GV galaxies (i.e., global sSFR below $10^{-10.5}$yr$^{-1}$). The black one-to-one line in each panel corresponds to equal contribution, above (below) which is \fgas~(SFE) driven. The background color denotes the classified gas quenching mode (orange: SFE driven; blue: \fgas~driven; green: mixed). The percentages of spaxels lying above and below the one-to-one line are indicated on either side of the dividing line. \label{fig:dSFE-dfgas}}
\end{figure*}

\subsection{Gas Mass, Stellar Mass, and Star Formation Rate Measurements}
In this work, all the physical quantities, including the H$_{2}$ mass surface density (\sigh2), stellar mass surface density (\sigsm), and star formation rate surface density (\sigsfr), are computed following the procedures described in \citet{lin20}. Here we summarize the key steps below. For a given spaxel, \sigh2~is computed from the CO luminosity by adopting a constant CO-to-\h2~ conversion factor ($\alpha_{\mathrm{CO}}$) of 4.35 \Msolar (K km s$^{-1}$ pc$^{2}$)$^{-1}$ \citep[e.g.,][]{bol13}. The impact of varying $\alpha_{\mathrm{CO}}$ to our results will be addressed in \S4.1. The stellar populations and ionized gas properties utilized in this work are taken from the MaNGA DR15 PIPE3D \citep{san16a,san16b} value-added products \citep{san18}, which contain both global properties (e.g., global SFR and global \sm~) and the spaxel-based measurements, such as \sigsm~ and emission-line fluxes. Corrections of dust extinction to emission line measurements are performed using the Balmer decrement computed on a spaxel basis assuming an intrinsic \ha/\hb = 2.86 and a Milky Way extinction curve with R$_v$ = 3.1 \citep{car89}. The extinction corrected \ha~flux is then converted to the SFR following the prescription given by Kennicutt (1998) with a Salpeter IMF. Since the global SFR and \sm~ values from PIPE3D are derived using only the spaxels within the MaNGA bundle coverage, we compute the global \Mh2~by summing the measurements over the same area for a given galaxy. The global SFE and global \fgas~are then calculated as SFR/\Mh2~ and \Mh2/\sm, respectively. To compute \sigsm, \sigh2, and \sigsfr, we normalize the stellar mass, gas mass, and SFR derived for each spaxel to the physical area of one spaxel corrected for inclination, taken from NASA Sloan Atlas (NSA) \footnote{https://www.sdss4.org/dr17/manga/manga-target-selection/nsa/}.

It should be noted that the SFR measurements used in this work are derived from extinction-corrected \ha~ luminosities, which may be contaminated by ionizing sources other than star formation, such as AGN, post-AGB stars, diffuse ionized gas (DIG), shocks, etc. \citep[e.g.,][]{sta08,sin13,smi22}, and should therefore be regarded as upper limits in regions not dominated by star formation \citep{sar10,yan12,sin13,hsi17,bel17,can19,lac20,ell21b}. To assess the impact of including AGN-like spaxels in our analysis, we use the standard Baldwin–Phillips–Terlevich (BPT) diagnostic diagrams \citep{bal81} to identify LIER and Seyfert spaxels, following dividing curves from the literature \citep[e.g.,][]{kew01,kau03,cid10}. We first determine whether a galaxy contains more than 10 BPT-classified AGN-like spaxels with \ha~ equivalent width EW(\ha) $>$ 3 \AA~ within the central 0.3 effective radius; this EW requirement helps remove emission powered by post-AGB stars \citep[e.g.,][]{cid11,lac20}. Using the \stwo-based diagnostics, we find that 0 (5) galaxies host central Seyfert (LIER)-like emission\footnote{The numbers become 0 (1) when using the \ntwo-based diagnostics.}. Although the number of LIER hosts is small, we mitigate the influence of LIER contamination by estimating, for each LINER-classified spaxel, the fractional contributions from star formation and LINER emission, denoted as $f_{SF}$ and $f_{L}$, respectively, by following Eq. (1) in \citet{bel18}:

\begin{equation}
f_{\mathrm{SF}}+f_{\mathrm{L}} = 1, 
\frac{\stwo}{\hamath}  = f_{\mathrm{SF}}\left(\frac{\stwo}{\hamath}\right)_{\mathrm{SF}} +  f_{\mathrm{L}}\left(\frac{\stwo}{\hamath}\right)_{\mathrm{L}}\,,
\end{equation}

where $f_{\mathrm{SF}}(\stwo/\hamath)_{\mathrm{SF}} = 0.4$ and $f_{\mathrm{L}}(\stwo/\hamath)_{\mathrm{L}} = 1$ \citep{bel16}. The upper and lower limits of $f_{\mathrm{SF}}$ are set to 1 and 0, respectively.

\subsection{Star Formation History}

To assess which driving factor affects the quenching timescales, we perform spectral fitting to reconstruct the star-formation histories (SFHs) of galaxies using MaNGA spectra coadded within the IFU coverage. Since one of our goals in the later analyses is to associate the star formation history with the gas properties of individual galaxies, the spectral fitting is performed on a per-galaxy basis.

The quenching timescale has been defined in a variety of different ways in the literature \citep[e.g.,][]{wet13,wri19,oma21,vis25} and should not be confused with others when making comparisons. For example, in studies using simulations for probing the environmental quenching effects, the quenching timescale is conveniently defined as the time between the epoch since the galaxy's first infall to dark matter halos and the time the star formation is quenched, i.e., below a certain sSFR threshold \citep[][]{wet13,oma21}. This can be in general longer compared to the star formation fading time, commonly referred to as the e-folding time over which SFR fades \citep{wet13}. Alternatively, the quenching timescale can be defined as the time between two thresholds in sSFR, as adopted by \citet[e.g.,][]{wal22}, who probed the quenching timescale using the IllustrisTNG simulation. Observationally, there is also diversity in the definitions of the quenching timescale. A common way to parameterize SFHs in the literature is based on a constant SFR for a certain period followed by an exponentially declining SFR with a variable e-folding time, which is then referred as quenching timescale \citep[e.g.,][]{sch14,nog19}. Other studies refer it to as the lookback time between a galaxy leaving the star-forming phase and entering the quiescent population \citep[e.g.,][]{tac22,bra23} or as the time after the Big Bang at which a galaxy is classified as quiescent \citep{ham23}. In addition, the change in the stellar metallicities has also been frequently used to indirectly disentangle between the fast and slow quenching \citep[e.g.,][]{pen15,tru20,leu24}. 

In this work, we model the star formation history (SFH) by performing a double-power law parametrization on integrated spectra, enabling a direct definition of the quenching timescale from the fit:  

\begin{equation}\label{eq:sfh}
	SFR(t) \propto \left[  \left(\frac{t}{\tau}\right)^\alpha + \left(\frac{t}{\tau}\right)^{-\beta} \right]^{-1},
\end{equation}
where $\tau$ is related to time of the peak in the SFH, while $\alpha$ and $\beta$ describe the declining and rising slope of the SFH, respectively. Here, $t$ represents cosmic time, measured from the beginning of the Universe. This double-power-law SFH model allows a more flexible parameterization of SFH than other approaches, such as exponentially declining function, log-normal, or delayed-$\tau$ models \citep{die17}. The characteristic rise and subsequent decline of the SFR is found to reproduce the SFHs of galaxies well, both in observational data \citep{gla13} and in simulations \citep{beh13,die17}, across a wide range of redshifts and galaxy populations. Thus, while this work focuses on the GV population, this general form is also applicable to typical MS galaxies.

For each galaxy, we first coadd the spectra from each spaxels to generate a 1D spectrum integrated over the entire IFU field of view. Spaxels with a signal-to-noise ratio ($S/N$) below 1 are excluded from the summation, as their contribution is predominantly noise, which would degrade the final coadd spectrum. Adopting a more stringent S/N threshold of up to 3 has a negligible impact on the results \citep{wu21}. The MaNGA Data Reduction Pipeline \citep[DRP;][]{law16} provides flags that indicate the quality of each pixel. Pixels marked with the DONOTUSE flag in the MaNGA quality bitmask (MANGA\_DRP3PIXMASK \footnote{https://sdss-mangadap.readthedocs.io/en/latest/metadatamodel.html}) should not be used for science. We replace the values of pixels flagged as DONOTUSE by values interpolated from neighboring pixels when summing up all the spectra from the spaxels considered \footnote{If a spaxel is entirely flagged as \texttt{DONOTUSE}, it will be excluded from the summation when generating the coadded spectrum.}.

We use \texttt{BAGPIPES} \citep{car18} to perform a Bayesian analysis to fit the integrated spectra. Stellar continuum templates are generated based on the updated \citet{bru03} stellar population synthesis model \citep{che16}, using the E-MILES stellar spectral library \citep{fal11,vaz16}. Dust attenuation is modeled with a power-law attenuation curve, $A_\lambda \propto \lambda^n$ \citep{cha00}, where the exponent $n$ is allowed to vary between 0.3 and 1.5. The total $V$-band attenuation, $A_v$, ranges from 0 and 4, with young stars ($t < 10^7$ yr) experiencing twice the attenuation of older stars. Emission lines are modeled self-consistently, with a ionization parameter constrained within $-3.5 < logU < -2$. We allow a wide range of ionization parameters to adopt different kinds of ionizing sources. Additionally, stellar metallicity and stellar velocity dispersions are treated as free parameters in the fitting. Redshifts are allowed to vary within a narrow range of $\pm0.005$ around the values provided in the MaNGA Data Analysis Pipeline \citep[DAP;][]{bel19,wes19} catalog to account for minor systematic uncertainties. The fitting is performed over the rest-frame wavelength range of 3650\AA--7000\AA, excluding regions affected by strong sky emission lines to mitigate potential influence from imperfect sky subtraction.

We define the quenching timescale, \taudecay, as the time required for the SFR to decline from its peak to 1/e ($\sim$37\%) of that value, i.e., the e-folding time that parametrizes the exponential decay. This definition emphasizes the duration over which the quenching process actively takes effect, regardless of the strength of the initial SFR or sSFR. Here, 'quenching' is used in a broader sense, encompassing both the gradual decline of star formation driven by the cosmic decrease in gas supply and the rapid suppression of star formation through more abrupt processes. Of the 71 galaxies in our sample, 24 have SFRs that do not fall below 1/e of their peak values within 10 Gyr of the present day and are therefore excluded from the subsequent analyses. All but one of the excluded galaxies belong to the MS population; therefore, this selection has a negligible impact on the results for GV galaxies presented in this work.
The best-fit parameters and the resulting derivation of \taudecay~ are given in Table \ref{tab:SFH}. The uncertainties of each parameter are computed using 16$^{th}$ and 84$^{th}$ percentiles relative to the 50$^{th}$ percentile of the posteriors.  Figure~\ref{fig:sfh} presents an example of the spectral fit and the corresponding SFH reconstruction. Similar figures for the full sample are given in Appendix A. 

\begin{figure}
\includegraphics[scale=0.32]{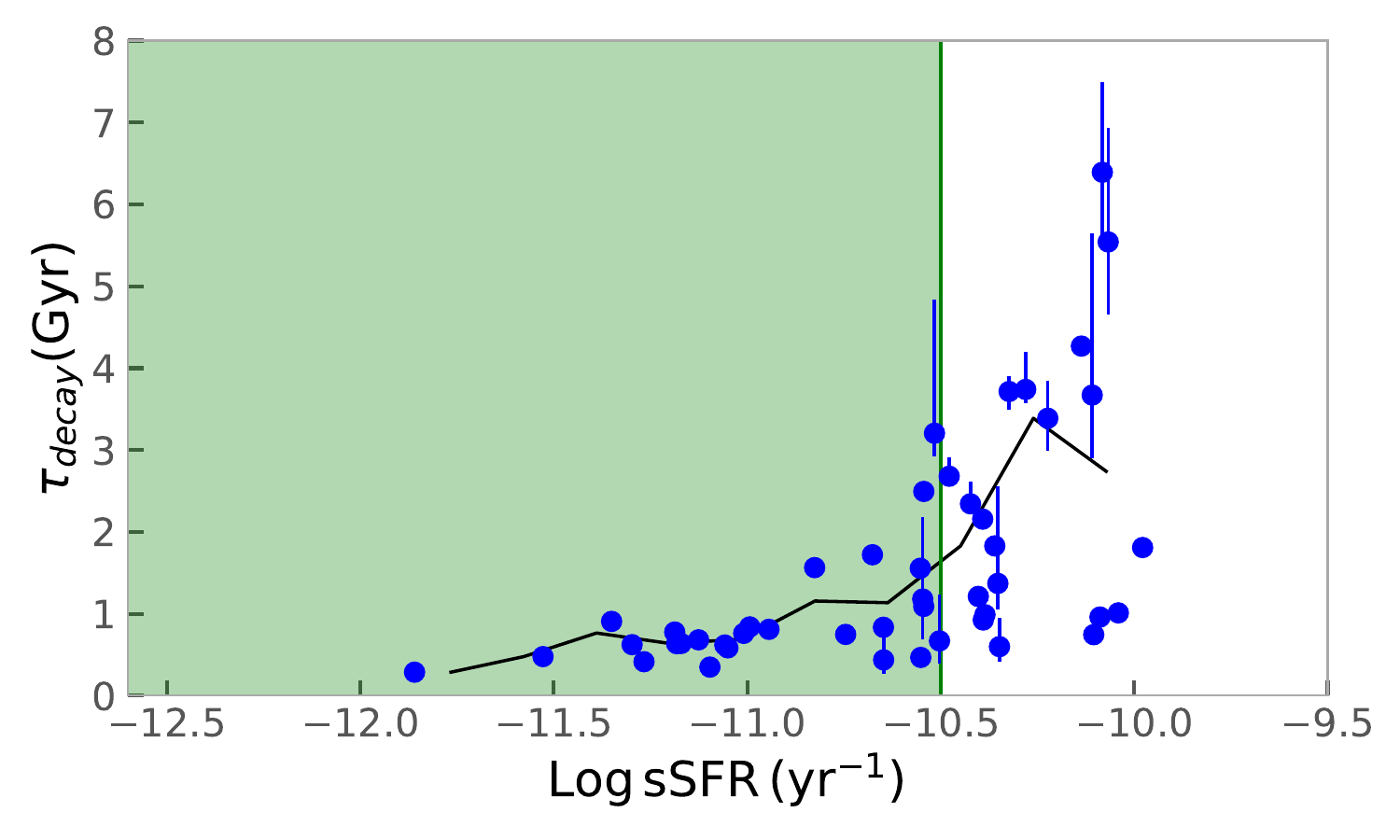}
\caption{Quenching timescale (\taudecay) as a function of global sSFR. The black curve represents the median \taudecay~values in bins of Log(sSFR). Uncertainties smaller than the symbol size are not visible in the plot. Galaxies with sSFR $<$ $10^{-10.5}$yr$^{-1}$ (indicated by the green background) define the GV population analyzed in this study. \label{fig:tau-ssfr}}
\end{figure}

\begin{figure*}
\includegraphics[scale=0.35]{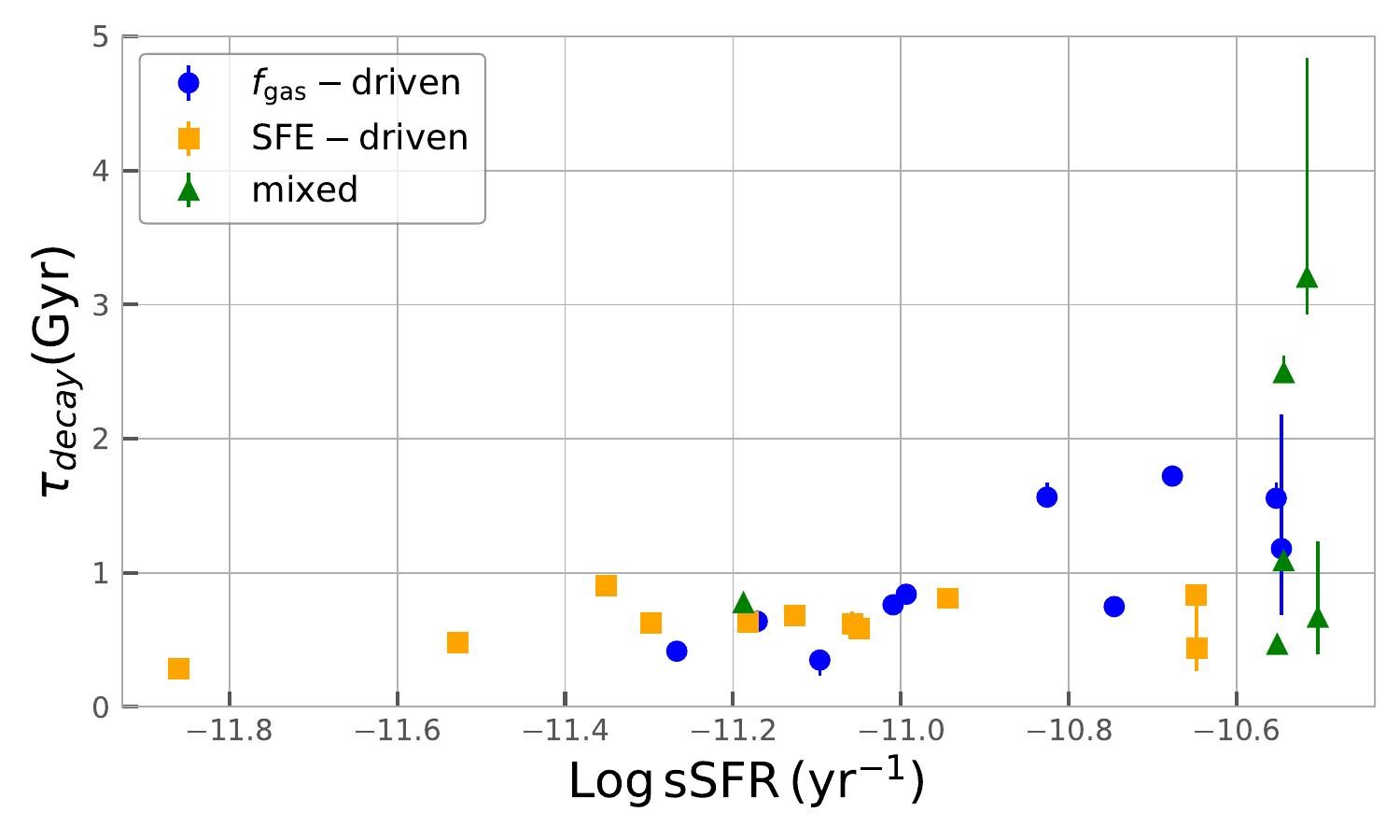}
\includegraphics[scale=0.35]{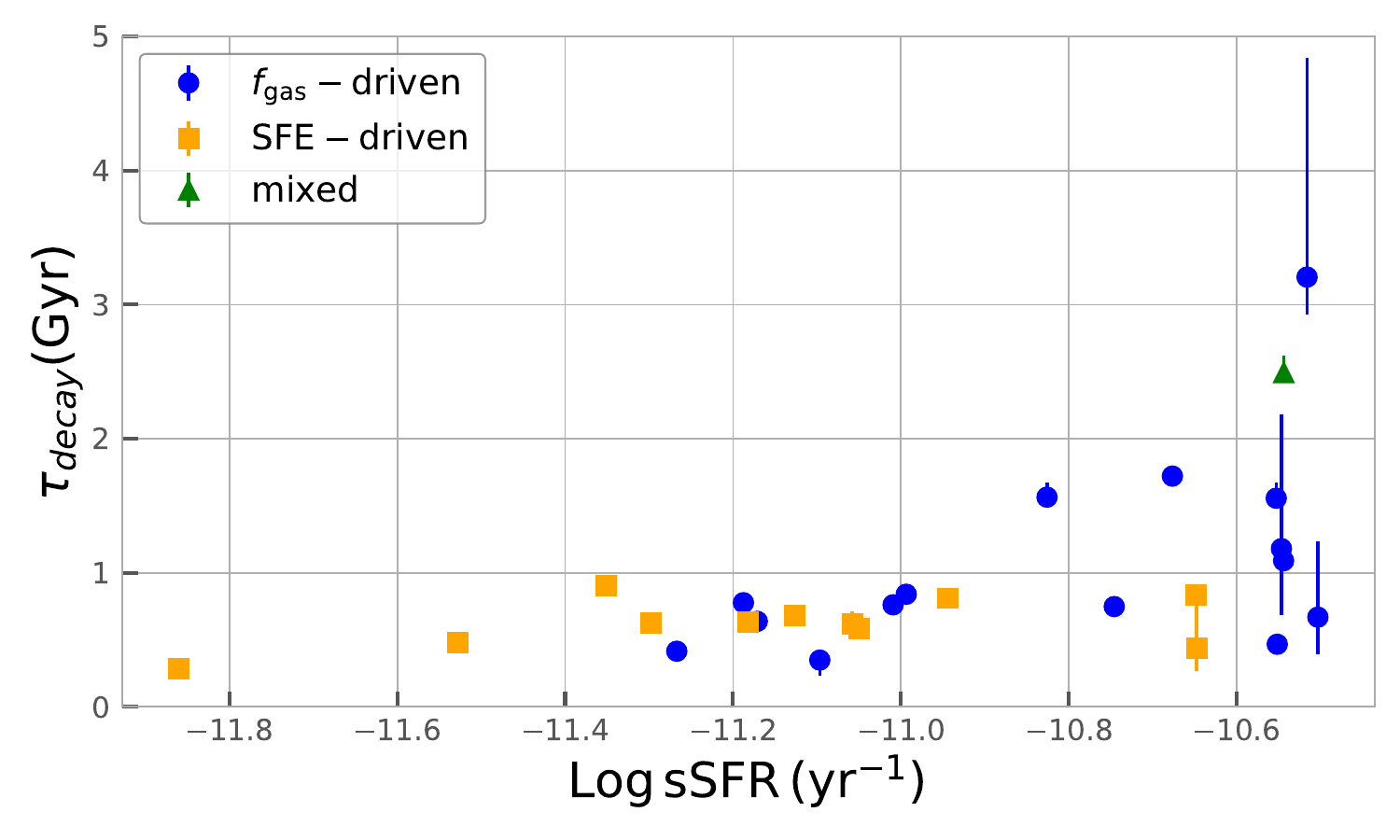}
\caption{Quenching timescale (\taudecay) as a function of global sSFR but for GV galaxies (i.e., sSFR below $10^{-10.5}$yr$^{-1}$) only. Colors denote different gas quenching modes (blue: gas-driven; orange: SFE-driven, green: mixed). Uncertainties smaller than the symbol size are not visible in the plot. Left: \Mh2~ is computed using a fixed  $\alpha_{\mathrm{CO}}$. Right: \Mh2~is computed using a metallicity-dependent  $\alpha_{\mathrm{CO}}$.\label{fig:tau-ssfr_qmode}}
\end{figure*}

\subsection{Classification of gas quenching modes}
One of our goals is to compare the relative contributions of SFE and \fgas~ to the suppression of sSFR in GV galaxies. Since previous studies have shown significant diversity in the resolved properties of galaxies even on the individual galaxy basis \citep{lin20,ell21a}, we opt for a spaxel-based approach rather than relying on global offsets in SFE and \fgas~relative to the main sequence. However, as illustrated in subsequent analyses, the spaxel-based classification of gas quenching modes is in good agreement with global SFE and \fgas~measurements. We will return to this point in \S3.2.2.

We follow the methodology described in \citet{tho22} to categorize the primary physical mode driving star formation based on resolved properties for a given galaxy. This method was originally designed to infer modes of enhanced star formation; here we modify it for the study of gas quenching modes. The main difference is that while \citet{tho22} focused on regions with enhanced star formation, we instead analyze regions where star formation is suppressed. For this analysis, we focus on 28 GV galaxies selected in \S2.1.

To quantify the relative contribution of SFE and \fgas~in suppressing star formation, we measure the offsets in both parameters, denoted as $\Delta$SFE and $\Delta$\fgas, respectively,  for each spaxel relative to the reference values derived from the best-fit of three resolved scaling relations of star-forming spaxels in MS galaxies in \citet{lin22}.  The fittings were performed using the orthogonal distance regression (ODR) method, and the fitting parameters are listed in Table 1 of \citet{lin22}.
For this analysis, we require both CO and \ha~to be detected with S/N $>$ 3 and \sigsm~ $>$ $10^{7.0}$ \Msolar kpc$^{-2}$. Spaxels that fail to meet these criteria are discarded. An obvious caveat is that this selection avoids spaxels that lack either \ha~ or CO measurements, which in any case cannot be used to robustly quantify SFE and \fgas~and potentially bias against \fgas~quenching.

We first identify spaxels with negative $\Delta$sSFR within the area enclosed by 1.5 \Re, which typically corresponds to about 60\% of the area covered by the MaNGA bundles.
Figure \ref{fig:dSFE-dfgas} presents the  $\Delta$SFE versus $\Delta$\fgas~diagram for each green valley galaxy, with the black line indicating the one-to-one relation where SFE and \fgas~contribute equally to the suppression of star formation. Galaxies are classified into quenching modes based on the following criteria:\\

\begin{itemize}
\item
If more than 60\% of the spaxels exhibit $\Delta$SFE $<$ $\Delta$\fgas~ (i.e., below the one-to-one line), the galaxy is classified as undergoing SFE-driven quenching.
\item If more than 60\% of the spaxels show $\Delta$SFE $>$ $\Delta$\fgas~(i.e., above the one-to-one line), the galaxy is classified as undergoing \fgas-driven quenching.
\item Galaxies that do not meet either criterion are categorized as experiencing a mixture of gas quenching modes. 
\end{itemize}

Similar to \citet{tho22}, this classification method assigns equal weight to all spaxels, regardless of their star formation strength. As a result, our analysis emphasizes the dominant mechanism across the majority of a galaxy’s regions, rather than the mechanism responsible for the areas with the most suppressed star formation. In most cases, suppression in both SFE and \fgas~is observed. The fraction of spaxels on either side of this relation and the corresponding quenching mode for individual galaxies are summarized in Table \ref{tab:sample}.

\section{RESULTS}

\subsection{Gas Quenching Modes}

Following the methodology described in \S2.4, Following the methodology described in §2.4, we classify 10 galaxies as SFE-driven ($N_{f_{\rm gas}}$ = 10), 11 as \fgas-driven ($N_{f_{\rm gas}}$ = 11), and 7 as mix-mode quenching ($N_{mix}$ = 7). The fractions of galaxies under these three quenching conditions are calculated as follows: $F_{SFE} = N_{SFE}/N_{GV}$, $F_{f_{\rm gas}} = N_{f_{\rm gas}}/N_{GV}$, and $F_{mix} = N_{mix}/N_{GV}$, where $N_{GV}$ corresponds to the number of GV galaxies. We find $F_{SFE}$, $F_{f_{\rm gas}}$, and $F_{mix}$ to be 35.7$\pm$13.2\%, 39.3$\pm$14.0\%, and 25.0$\pm$10.6\%, respectively.

Overall, our results using a fixed $\alpha_{\mathrm{CO}}$ indicate significant galaxy-to-galaxy variation in the gas quenching modes among GV galaxies, and in most cases, even within individual galaxies themselves, as also noted by \citet{pan24}. In \S4.1, we examine how this conclusion may change when adopting a variable $\alpha_{\mathrm{CO}}$.

\subsection{Quenching Timescale}

\subsubsection{\taudecay~vs. sSFR}
After investigating the gas quenching modes, we now turn to explore the quenching timescale of these galaxies. We start by showing \taudecay~as a function of global sSFR for the 47 galaxies with \taudecay~measured in Figure \ref{fig:tau-ssfr}.  
A clear trend is observed, showing an increase \taudecay~ with rising sSFR of galaxies. This trend is somewhat expected, as star-forming galaxies, by definition, are still actively forming stars, either because they have not yet experienced significant quenching effects or they undergo a more gradual suppression of star formation and are still in the early phase of quenching. This trend is not biased by the exclusion of the 24 galaxies with long \taudecay, as those galaxies are largely dominated by the MS population (see \S2.3) and would in fact amplify the trend observed here. In contrast, galaxies with low sSFR are generally in an advanced stage of quenching, which is more likely to occur in systems with relatively short quenching timescales. In the subsequent analyses, we only focus on GV galaxies (i.e., sSFR below $10^{-10.5}$yr$^{-1}$), in which the gas quenching modes are classified as described in \S3.1.

In the left panel of Figure \ref{fig:tau-ssfr_qmode}, we display \taudecay~as a function of global sSFR but for GV galaxies only, color-coded according to the gas quenching modes (blue: \fgas-driven; orange: SFE-driven; green: mixed). The quenching mode changes from multiple quenching modes to SFE-driven quenching only as galaxies move toward low sSFR regime, albeit our sample size is still limited. Similar results were also reported in the EDGE–CALIFA survey, where a gradual decline in \fgas~is observed from main-sequence to quiescent populations \citep{col20,vil24}. However, these studies also noted that a reduction in SFE is required for galaxies to become fully quenched.

\subsubsection{\taudecay~as a function of global \fgas~and SFE}
\begin{figure*}[tbh]

\includegraphics[scale=0.7]{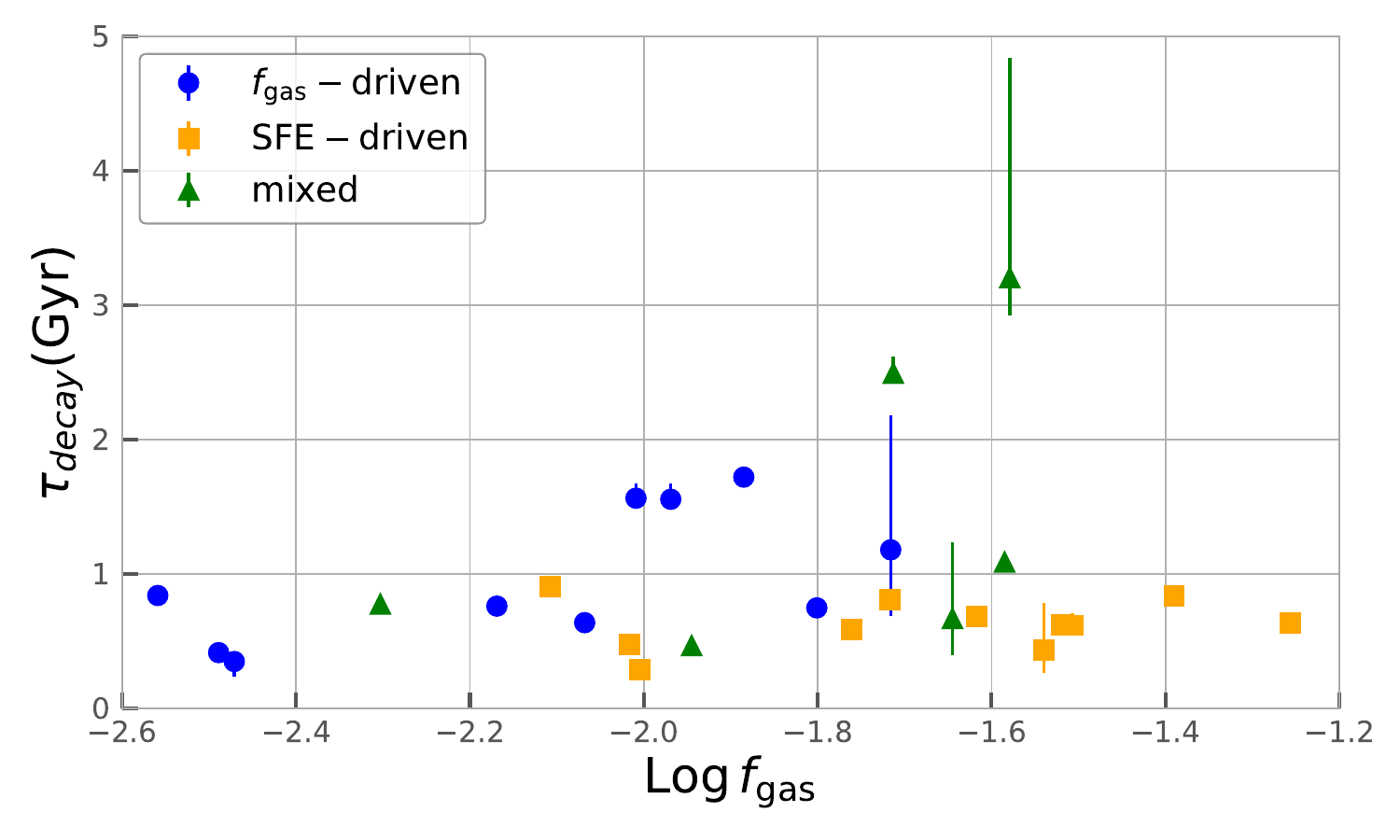}
\caption{Quenching timescale (\taudecay) as a function of global \fgas. Only galaxies defined as GV galaxies are shown with colors denoting different gas quenching modes (blue: gas-driven; orange: SFE-driven, green: mixed). Uncertainties smaller than the symbol size are not visible in the plot. \label{fig:tau-fgas}}

\end{figure*}

\begin{figure*}[tbh]

\includegraphics[scale=0.7]{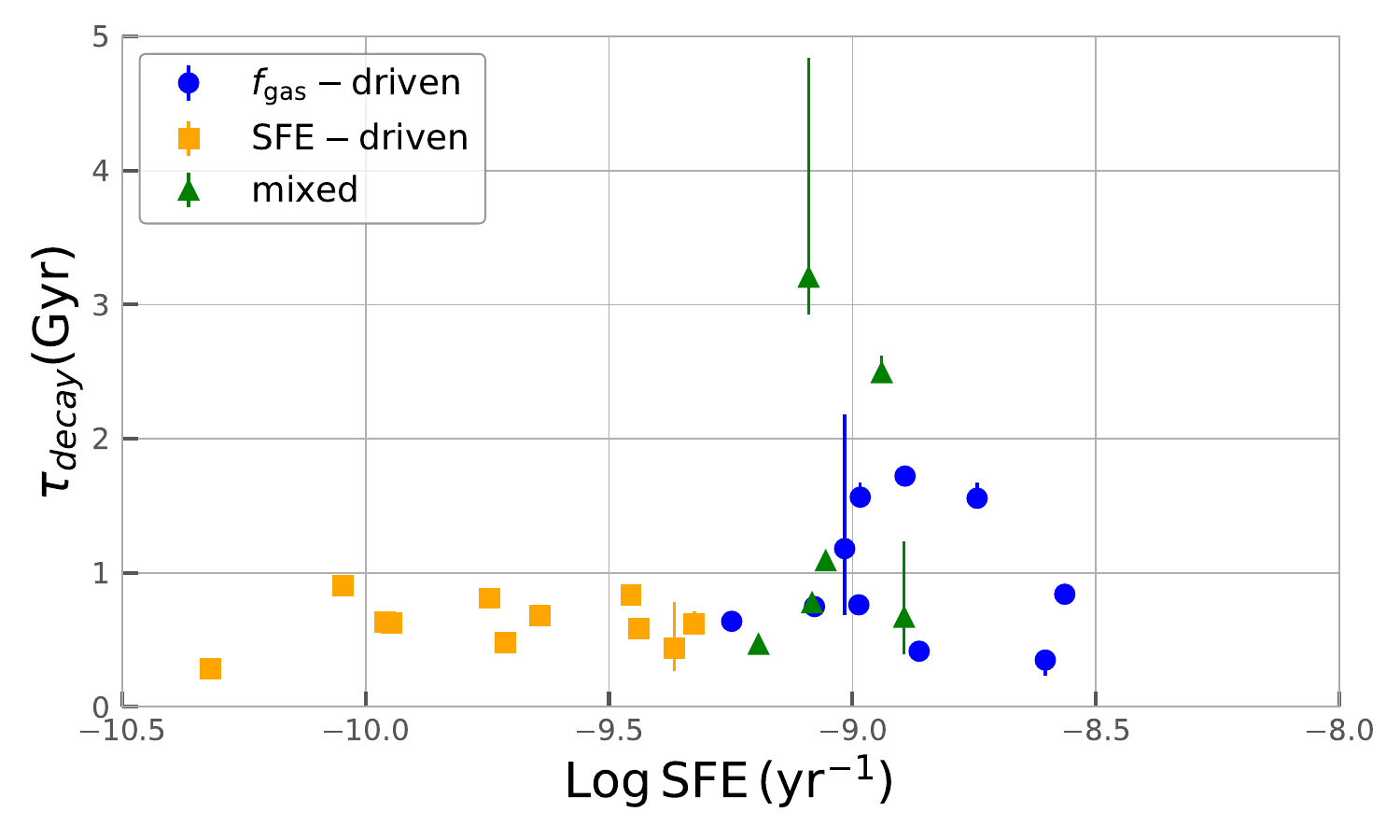}
\caption{Quenching timescale (\taudecay) as a function of global SFE. Only galaxies defined as GV galaxies are shown with colors denoting different gas quenching modes (blue: gas-driven; orange: SFE-driven, green: mixed). Uncertainties smaller than the symbol size are not visible in the plot. \label{fig:tau-sfe}}

\end{figure*}

In Figure \ref{fig:tau-fgas}, we plot \taudecay~ as a function of global \fgas~ for GV galaxies, color-coded by their corresponding gas quenching modes (blue: gas-driven; orange: SFE-driven, green: mixed). No obvious correlation between \taudecay~and global \fgas~is revealed. The majority (75\%) of GV galaxies have \taudecay~values below 1 Gyr. The short quenching timescale disfavors strangulation (or starvation) being the major quenching mode in these GV galaxies, as the strangulation process is expected to operate over long timescales, typically 2–4 Gyr \citep[e.g.,][]{pen15,bax25}, depending on gas consumption efficiency and initial gas content. It is also evident that systems undergoing SFE-driven quenching do not necessarily exhibit a deficit in molecular gas.

\begin{figure}[tbh]
\includegraphics[angle=0,width=0.47\textwidth]{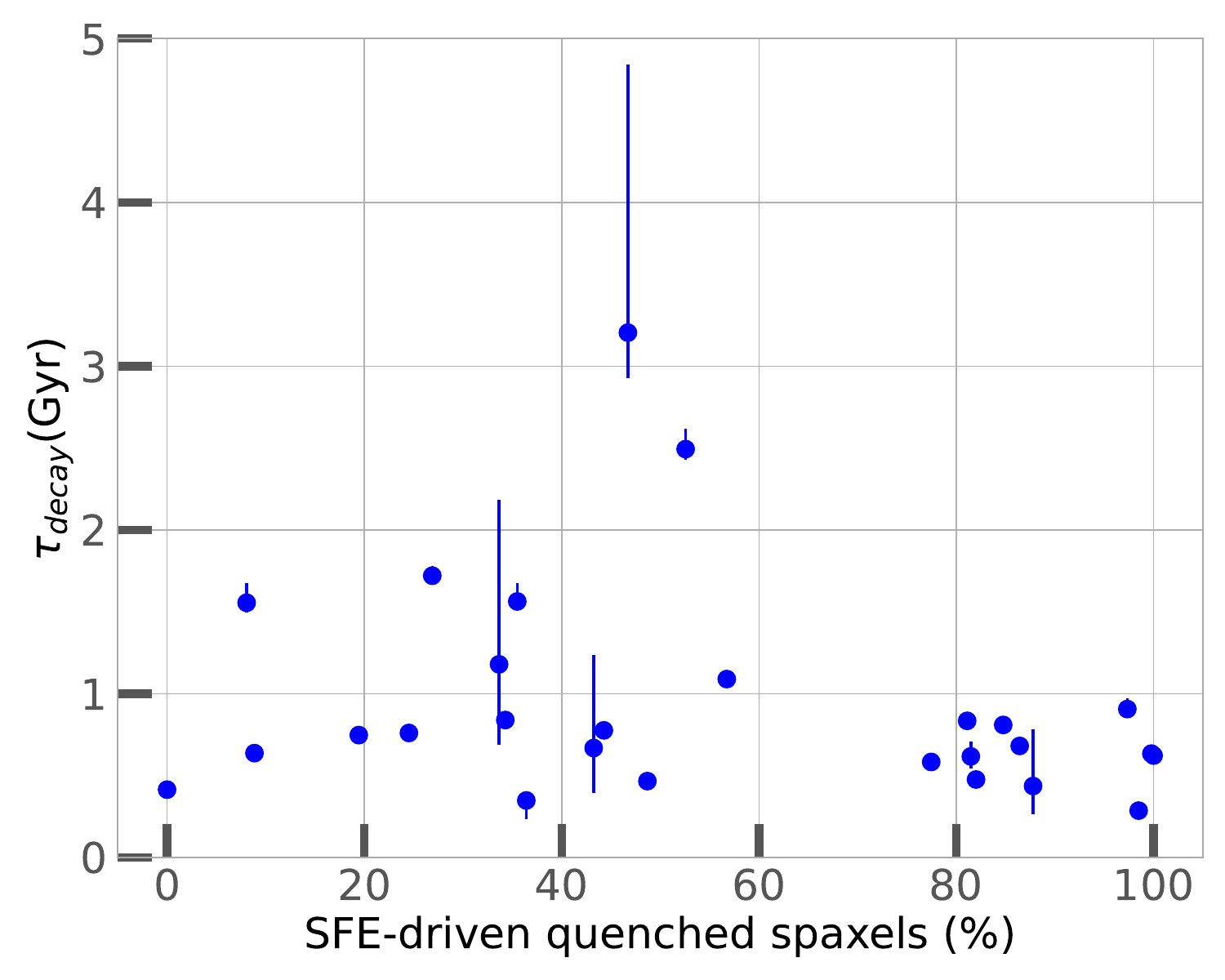}
\caption{\taudecay~vs. the fraction of quenched spaxels due to SFE driven. Uncertainties smaller than the symbol size are not visible in the plot. \label{fig:tau-sfe_percent}}
\end{figure}

Next, we examine the dependence of \taudecay~ on global SFE, as shown in Figure \ref{fig:tau-sfe}. Significant variation in \taudecay~ is observed only in galaxies with SFE above a certain threshold (approximately $10^{-9.2}\mathrm{yr^{-1}}$); below this value, \taudecay~ tends to remain consistently below 1 Gyr. An alternative way to explore this relationship is to consider \taudecay~ as a function of the fraction of SFE-driven quenched spaxels on a galaxy-by-galaxy basis (Figure \ref{fig:tau-sfe_percent}). We find that when the fraction of SFE-driven spaxels exceeds 60\%—indicating that the galaxy is predominantly in the SFE-driven quenching mode—\taudecay~ consistently falls below 1 Gyr. This suggests that low SFE is a sufficient condition for rapid quenching.

Another thing to note is that galaxies classified as \fgas-driven quenching tend to have lower global \fgas~ compared to those classified as SFE-quenching (Figure 6). Similarly, galaxies classified as SFE-driven quenching tend to have low global SFE compared to those classified as \fgas-quenching (Figure 7). This suggests that the spatially-resolved classification in gas quenching modes is in good agreement with the global strength of \fgas~ and SFE.

\subsubsection{\taudecay~ in various quenching modes}

\begin{figure*}[tbh]
\includegraphics[angle=0,width=0.45\textwidth]{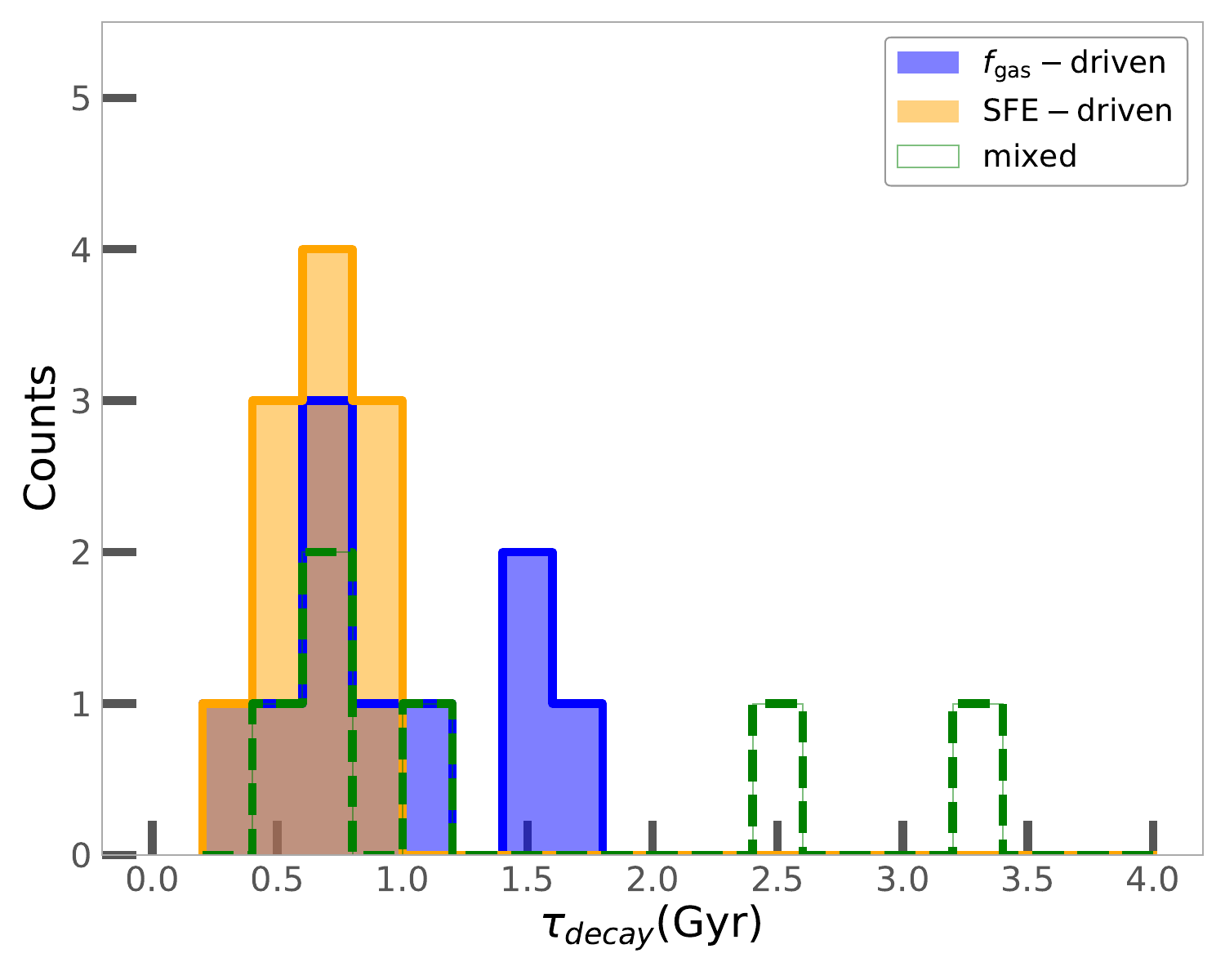}
\includegraphics[angle=0,width=0.45\textwidth]{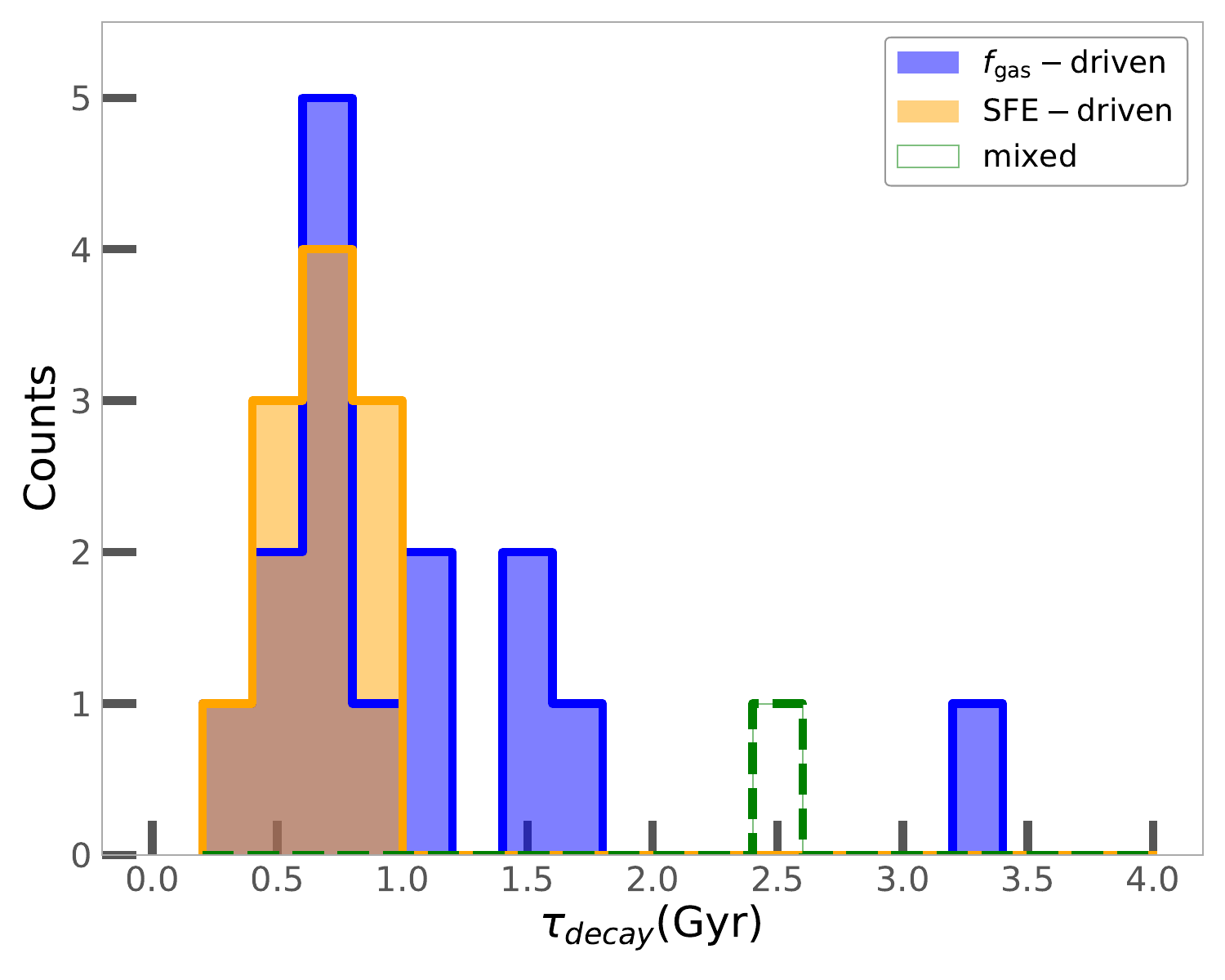}
\caption{Histogram distributions of \taudecay~for various quenching modes. Blue-shaded regions are \fgas-driven, orange-shaded regions are SFE-driven, and green dotted regions indicate mixed behavior. Left: \Mh2~ is computed using a fixed  $\alpha_{\mathrm{CO}}$. Right: \Mh2~is computed using a metallicity-dependent  $\alpha_{\mathrm{CO}}$.   \label{fig:his_tau}}
\end{figure*}

To further investigate the connection between the quenching timescale and the gas quenching mode, we present the distribution of \taudecay~ of the three molecular gas quenching modes in the left panel of Figure \ref{fig:his_tau}. As already mentioned in \S3.2.2, most of the GV galaxies exhibit \taudecay~less than 1 Gyr, i.e., fast quenching mode, regardless of whether they are SFE or \fgas~driven quenching. The short timescale obtained from our GV sample is consistent with other studies of green valley galaxies in the literature, which attribute the scarcity of galaxies in the green valley to a relatively fast transition from the star-forming to the quiescent population \citep[e.g.,][]{sch14, bre18}.

There is a slight indication that the mixed mode exhibits a tail extending toward longer \taudecay, followed by the \fgas-driven mode, whereas the SFE-driven mode tends to have shorter \taudecay. However, this trend is marginal given the small sample size for each molecular gas quenching mode, especially for galaxies with the mixed quenching mode.

\section{Discussion}
\subsection{The impact of $\alpha_{\mathrm{CO}}$}
Our analyses thus far are based on \h2~mass computed using a constant $\alpha_{\mathrm{CO}}$ = 4.35 \Msolar (K km s$^{-1}$ pc$^{2}$)$^{-1}$. However, it has been shown that $\alpha_{\mathrm{CO}}$ can vary both globally and across galactic scales, particularly at the galaxy centers \citep[e.g.][]{san13}. Several prescriptions for this variation have been proposed in the literature, taking into account factors such as metallicity, distance from the galactic center, velocity dispersion, and gas surface density \citep[e.g.,][]{nar12, sun20, ten22, ten23, chi25, ler25}. Because these prescriptions involve different assumptions and depend on the physical scale of measurement, not all of them can be directly applied to our dataset. To assess the potential impact of a varying $\alpha_{\mathrm{CO}}$ on the inferred gas quenching modes, we adopt a metallicity-dependent $\alpha_{\mathrm{CO}}$ in this section, following Equation (4) of \citet{sun20}:

\begin{equation}\label{eq: 2d}
\alpha_{\mathrm{CO}} = 4.35 (Z/Z_{\odot})^{-1.6}~ \rm M_{\odot} ~pc^{-2}(K~km~s^{-1})^{-1} ,
\end{equation}
where Z is the (linear) gas phase abundance and Z$_{\odot}$ is the solar value that corresponds to 12+log(O/H) = 8.69. The gas phase metallicity in log can be calculated through the O3N2 calibrator \citep{pet04}: 
\begin{equation}
12 + \mathrm{log (O/H)} = 8.73 - 0.32 \times O3N2.
\end{equation}

Metallicity is computed only for spaxels classified as star-forming according to the selection criteria described in §2.2, which account for approximately 54\% of the spaxels in our sample. For non–star-forming regions, we adopt a conservative approach by assigning the median metallicity of the star-forming spaxels within the same galaxy or the solar metallicity when no star-forming spaxels are present in a given galaxy. We recomputed the spaxel-based $\Sigma_{\mathrm{H_2}}$, along with the corresponding $f_{\mathrm{gas}}$ and SFE, using the metallicity-dependent conversion factor. Five galaxies originally classified as mixed mode under a fixed $\alpha_{\mathrm{CO}}$ are reclassified as $f_{\mathrm{gas}}$-driven when adopting the metallicity-dependent $\alpha_{\mathrm{CO}}$. The number of $f_{\mathrm{gas}}$-driven galaxies hence increases from 11 to 16, while the numbers of mixed-mode galaxies decrease from 6 to 1. On the other hand, the number of SFE-driven galaxies remains unchanged. As a result, the relative contributions of the gas quenching modes become 57.1$\pm$17.9\%, 39.3$\pm$14.0\%, and 3.6$\pm$3.6\% for \fgas-driven, SFE-driven, and mixed modes, respectively. In other words, the fraction of \fgas-driven quenching slightly increases while the fraction of galaxies classified as mixed mode decreases when compared to the fixed $\alpha_{\mathrm{CO}}$. 
Because of this change, the tendency that mixed modes of GV galaxies tend to have greater sSFR and longer \taudecay~ no longer holds (see right panels of Figure \ref{fig:tau-ssfr_qmode} and Figure \ref{fig:his_tau}). On the other hand, the main conclusions that SFE-driven GVs tend to have shorter \taudecay~than \fgas-driven or mixed-mode GVs, and that multiple quenching modes are present in the GV sample remain solid.

\subsection{The connection between \taudecay~and gas quenching modes}

Our results reveal only a moderate correlation between the quenching timescale and the gas quenching mode. While all SFE-driven GV galaxies exhibit $\tau_{\mathrm{decay}}$ shorter than 1 Gyr, the $f_{\mathrm{gas}}$-driven cases span a much broader range of timescales—up to 3.5 Gyr, depending on the adopted $\alpha_{\mathrm{CO}}$. As described in Section 1, $f_{\mathrm{gas}}$-driven quenching can be associated with both rapid and slow quenching \citep{di05,kav11,muz14,her19,bax25}. Environmental mechanisms beyond ram-pressure stripping (which drives fast quenching), such as gravitational interactions or starvation, can reduce $f_{\mathrm{gas}}$ by removing or depleting the hot gas reservoir, thereby suppressing gas replenishment over extended timescales \citep{bax25}. The diverse $\tau_{\mathrm{decay}}$ values observed in $f_{\mathrm{gas}}$-driven GV galaxies thus suggest that multiple quenching processes may indeed be at play, varying from one galaxy to another. Investigating the large-scale environments of \fgas-driven galaxies using a larger sample, along with adopting a more stringent sample selection (e.g., restricting to only field galaxies or only interacting/cluster galaxies), can help disentangle whether the reduced gas fraction arises from intrinsic processes or environmental effects.

The consistently low $\tau_{\mathrm{decay}}$ observed in SFE-driven GV galaxies within our sample is intriguing. Several mechanisms have been suggested to drive low SFE, including morphological quenching \citep{mar09,kel19,oxl24}, radio-mode AGN feedback \citep{cro06,wei17,ven21}, and environmental effects associated with disturbed atomic gas, such as those observed in the VIRGO cluster \citep{vil22}. In the case of morphological quenching, the short $\tau_{\mathrm{decay}}$ may imply that the morphological transformation occurs on relatively short timescales, likely triggered by violent processes such as galaxy mergers or intense central starbursts. On the other hand, while radio AGN feedback is typically predicted by simulations to operate over longer timescales \citep{cro06,wei17,ven21}, the short $\tau_{\mathrm{decay}}$ seen in our SFE-driven sample may suggest that rapid quenching occurs earlier—before the onset of radio-mode AGN activity.

There are, however, several caveats and limitations in our analysis. First, it is plausible that some GV galaxies undergo multiple quenching episodes \citep{sch14}, which could obscure or complicate the relationship between gas properties and $\tau_{\mathrm{decay}}$. For instance, some studies have shown no significant differences between GV galaxies with and without AGN activity \citep[e.g.,][]{baz25}, suggesting that the cumulative impact of AGN feedback arises from a series of intermittent AGN episodes. The net effect, on average, is a reduction in star-formation activity, regardless of whether or not a galaxy is currently hosting an AGN. Another example is that if we assume that a short $\tau_{\mathrm{decay}}$ corresponds to SFE-driven quenching and long $\tau_{\mathrm{decay}}$ to $f_{\mathrm{gas}}$-driven quenching, a galaxy that first experiences rapid quenching via SFE followed by a slow $f_{\mathrm{gas}}$-driven event could present a misleading picture. In such a case, the current $f_{\mathrm{gas}}$ would appear low, while $\tau_{\mathrm{decay}}$ remains low, masking the underlying sequence of quenching processes.

Secondly, the star formation history (SFH) used in this work is based on a parametric approach using a double-power law (see \S2.3). Despite that it has been shown to produce the SFHs of observed and simulated galaxies reasonably well \citep{gla13,beh13,die17}, this simplified approach may fail to describe SFH of galaxies experiencing multiple episodes of starburst and/or quenching. On the other hand, it has been shown that stochastic star formation histories and non-parametric models can better predict and recover galaxy properties and SFH using simulations \citep{wan24}. Non-parametric treatments of SFH also provide more accurate descriptions of diverse galaxy populations and reduce biases associated with traditional SFH parametrizations \citep[e.g.,][]{ive14,iye19}. 
Nevertheless, this study represents an initial effort to investigate the connection between gas quenching modes and quenching timescales. Future work incorporating non-parametric SFH modeling will be crucial for testing the robustness of these results across a broader range of galaxy evolutionary histories.

\section{SUMMARY \label{sec:summary}}

In this work, we analyzed 71 galaxies with MaNGA integral field spectroscopy and matched-resolution ALMA \co~observations, selected from the ALMaQUEST \citep{lin20} and other samples with similar types of observations in the literature \citep{tho22,ott22}. For each galaxy, we compute the spaxel-based values of $\Delta$sSFR, $\Delta$\fgas~and $\Delta$SFE, and classify galaxies into \fgas-driven, SFE-driven, and mixed quenching modes, depending on the dominant mechanisms in regions with suppressed sSFR. We also derived quenching timescales by fitting a double-power-law model to the star formation histories inferred from the MaNGA integrated spectra. Our main results can be summarized as below:\\

1. We find a diversity of gas quenching modes among 28 GV galaxies, based on their molecular gas properties when using a fixed $\alpha_{\mathrm{CO}}$: 35.7$\pm$13.2\% are 
 \fgas-driven, 39.3$\pm$14.0\% are SFE-driven, and 25.0$\pm$10.6\% exhibit a mixed mode (Figure \ref{fig:dSFE-dfgas}). When adopting a metallicity-dependent $\alpha_{\mathrm{CO}}$, the fraction of \fgas-driven galaxies increases to 57.1$\pm$17.9\%, while the fraction of mixed modes decreases to 3.6$\pm$3.6\%; the fraction of SFE-driven galaxies remains unchanged. Overall, our results indicate that GV galaxies undergo quenching through multiple pathways.\\

2. We observe a trend of decreasing quenching timescale (\taudecay) with decreasing sSFR, suggesting that galaxies further from the main sequence tend to quench more rapidly (Figure \ref{fig:tau-ssfr}). When limited to GV galaxies only, the majority of galaxies (75\%) exhibit quenching timescales shorter than 1 Gyr, favoring fast quenching processes and disfavoring purely slow-quenching scenarios like starvation.\\

3. GV galaxies with high sSFR exhibit a variety of quenching modes, while those with the lowest sSFR are primarily classified as undergoing SFE-driven quenching (Figure \ref{fig:tau-ssfr_qmode}). This suggests that a reduced star formation efficiency (SFE) plays a crucial role in significantly suppressing star formation.\\

4. SFE-driven quenching is consistently associated with short quenching timescales regardless of the adopted  CO-to-\h2~ conversion factors. On the other hand, \fgas-driven quenching shows a broader range of \taudecay~(Figure \ref{fig:his_tau}), potentially reflecting a mix of fast and slow processes (e.g., AGN feedback and environmental gas removal).\\

 Our results suggest that SFE-driven quenching—such as that induced by morphological quenching or AGN feedback—can lead to a rapid suppression of star formation while significant molecular gas remains. In contrast, $f_{\mathrm{gas}}$-driven quenching may arise from a mix of rapid mechanisms like AGN feedback and more gradual processes such as gas depletion due to environmental effects. These findings highlight the importance of considering multiple dimensions, such as the efficiency of star formation, the availability of gas, as well as the timescale over which quenching occurs, for building up a comprehensive view. Nevertheless, our current sample remains limited, with only about 10 galaxies in each gas quenching category. The ongoing KILOGAS project (T. Davis et al., in prep.) will expand the green valley sample size by at least a factor of three, enabling a more robust characterization of the connection between gas properties and quenching timescales.
 \begin{acknowledgments}
We thank the anonymous referee for his/her helpful comments, which greatly improve the contents of this paper. 
This work is supported by the National Science and Technology Council of Taiwan under the grants NSTC 113-2112-M-001-006- and NSTC 114-2112-M-001-041-MY3. PFW thanks support by the National Science and Technology Council of Taiwan under grant 113-2112-M-002-027-MY2. HAP acknowledges support by the National Science and Technology Council of Taiwan under grant 110-2112-M-032-020-MY3.

The authors would like to thank the staff of the East-Asia and North-America ALMA ARCs for their support and continuous efforts in helping produce high-quality data products. This paper makes use of the following ALMA data:\\
ADS/JAO.ALMA\#2015.1.01225.S,\\  ADS/JAO.ALMA\#2017.1.01093.S,\\
ADS/JAO.ALMA\#2018.1.00541.S,\\
ADS/JAO.ALMA\#2018.1.00558.S.\\
ADS/JAO.ALMA\#2019.1.01178.S.\\
ALMA is a partnership of ESO (representing its member states), NSF (USA) and NINS (Japan), together with NRC (Canada), MOST and ASIAA (Taiwan), and KASI (Republic of Korea), in cooperation with the Republic of Chile. The Joint ALMA Observatory is operated by ESO, AUI/NRAO and NAOJ.

Funding for the Sloan Digital Sky Survey IV has been
provided by the Alfred P. Sloan Foundation, the U.S.
Department of Energy Office of Science, and the Participating Institutions. SDSS-IV acknowledges support
and resources from the Center for High-Performance
Computing at the University of Utah. The SDSS web
site is www.sdss.org. SDSS-IV is managed by the Astrophysical Research Consortium for the Participating
Institutions of the SDSS Collaboration including the
Brazilian Participation Group, the Carnegie Institution
for Science, Carnegie Mellon University, the Chilean
Participation Group, the French Participation Group,
Harvard-Smithsonian Center for Astrophysics, Instituto
de Astrof\'isica de Canarias, The Johns Hopkins University, Kavli Institute for the Physics and Mathematics of the Universe (IPMU) / University of Tokyo, Lawrence
Berkeley National Laboratory, Leibniz Institut f\"ur Astrophysik Potsdam (AIP), Max-Planck-Institut f\"ur Astronomie (MPIA Heidelberg), Max-Planck-Institut f\"ur
Astrophysik (MPA Garching), Max-Planck-Institut f\"ur
Extraterrestrische Physik (MPE), National Astronomical Observatory of China, New Mexico State University,
New York University, University of Notre Dame, Observat\'ario Nacional / MCTI, The Ohio State University,
Pennsylvania State University, Shanghai Astronomical
Observatory, United Kingdom Participation Group, Universidad Nacional Aut\'onoma de M\'exico, University of
Arizona, University of Colorado Boulder, University of
Oxford, University of Portsmouth, University of Utah,
University of Virginia, University of Washington, University of Wisconsin, Vanderbilt University, and Yale University. 
\end{acknowledgments}

\startlongtable
\begin{deluxetable*}{llccccccc}
	\tabletypesize{\scriptsize}
	\tablewidth{0pt}
	\tablecaption{Properties of 71 Galaxies Used in This Work \label{tab:sample}}
	\tablehead{
		\colhead{Plateifu} &
		\colhead{Sample$^{(a)}$} &
		\colhead{log$_{10}$(\sm/\Msolar)} &
		\colhead{log$_{10}$($\frac{\rm sSFR}{\rm yr^{-1}}$)} &
		\colhead{log$_{10}$($\frac{\rm SFE}{\rm yr^{-1}}$)} &
		\colhead{log$_{10}$\fgas}&
		\colhead{S$_{SFE}$$^{(b)}$}&
		\colhead{S$_{gas}$$^{(c)}$}&
		\colhead{Q$_{m}$$^{(d)}$}
	}
\startdata
7815-12705 & Lin+2020 & 10.83 & -10.42 & -8.87 & -1.1 & \nodata & \nodata & \nodata \\
 7977-3703 & Lin+2020 & 10.35 & -9.95 & -8.73 & -1.3 & \nodata & \nodata & \nodata \\
 7977-3704 & Lin+2020 & 10.36 & -10.75 & -9.08 & -1.8 & 0.19 & 0.81 & $f_{gas}$ \\
 7977-9101 & Lin+2020 & 11.23 & -11.10 & -8.60 & -2.5 & 0.36 & 0.64 & $f_{gas}$ \\
 7977-12705 & Lin+2020 & 10.85 & -10.39 & -8.68 & -1.6 & \nodata & \nodata & \nodata \\
 8077-6104 & Lin+2020 & 10.73 & -10.08 & -8.83 & -1.2 & \nodata & \nodata & \nodata \\
 8077-9101 & Lin+2020 & 10.41 & -10.50 & -8.89 & -1.6 & 0.43 & 0.57 & mixed \\
 8078-6103 & Lin+2020 & 10.75 & -10.14 & -8.96 & -1.3 & \nodata & \nodata & \nodata \\
 8078-12701 & Lin+2020 & 10.95 & -10.54 & -9.06 & -1.6 & 0.57 & 0.43 & mixed \\
 8081-3704 & Lin+2020 & 10.56 & -9.59 & -8.19 & -1.3 & \nodata & \nodata & \nodata \\
 8081-6102 & Lin+2020 & 10.79 & -11.19 & -9.08 & -2.3 & 0.44 & 0.56 & mixed \\
 8081-9101 & Lin+2020 & 10.60 & -10.28 & -9.02 & -1.4 & \nodata & \nodata & \nodata \\
 8081-9102 & Lin+2020 & 10.66 & -10.42 & -8.99 & -1.6 & \nodata & \nodata & \nodata \\
 8081-12703 & Lin+2020 & 10.34 & -11.13 & -9.64 & -1.6 & 0.86 & 0.14 & SFE \\
 8082-6103 & Lin+2020 & 10.31 & -9.92 & -8.53 & -1.4 & \nodata & \nodata & \nodata \\
 8082-12701 & Lin+2020 & 10.48 & -10.36 & -8.74 & -1.7 & \nodata & \nodata & \nodata \\
 8082-12704 & Lin+2020 & 11.42 & -10.65 & -9.37 & -1.5 & 0.88 & 0.12 & SFE \\
 8083-6101 & Lin+2020 & 10.30 & -11.18 & -9.96 & -1.3 & 1.00 & 0.00 & SFE \\
 8083-9101 & Lin+2020 & 11.14 & -10.54 & -8.94 & -1.7 & 0.53 & 0.47 & mixed \\
 8083-12702 & Lin+2020 & 11.22 & -10.56 & -8.77 & -1.7 & 0.44 & 0.56 & mixed \\
 8084-3702 & Lin+2020 & 10.23 & -9.80 & -8.80 & -1.2 & \nodata & \nodata & \nodata \\
 8084-6103 & Lin+2020 & 10.51 & -10.65 & -9.45 & -1.4 & 0.81 & 0.19 & SFE \\
 8084-12705 & Lin+2020 & 10.45 & -10.48 & -8.84 & -1.8 & \nodata & \nodata & \nodata \\
 8086-9101 & Lin+2020 & 10.94 & -10.83 & -8.98 & -2.0 & 0.35 & 0.65 & $f_{gas}$ \\
 8155-6101 & Lin+2020 & 10.93 & -11.35 & -10.05 & -2.1 & 0.97 & 0.03 & SFE \\
 8155-6102 & Lin+2020 & 10.36 & -10.00 & -8.88 & -1.1 & \nodata & \nodata & \nodata \\
 8156-3701 & Lin+2020 & 10.52 & -9.65 & -8.11 & -1.4 & \nodata & \nodata & \nodata \\
 8241-3703 & Lin+2020 & 10.11 & -9.86 & -8.57 & -1.4 & \nodata & \nodata & \nodata \\
 8241-3704 & Lin+2020 & 11.00 & -9.79 & -8.78 & -1.0 & \nodata & \nodata & \nodata \\
 8450-6102 & Lin+2020 & 10.43 & -9.79 & -8.56 & -1.2 & \nodata & \nodata & \nodata \\
 8615-3703 & Lin+2020 & 10.19 & -9.79 & -8.81 & -1.1 & \nodata & \nodata & \nodata \\
 8615-9101 & Lin+2020 & 10.58 & -10.55 & -9.02 & -1.7 & 0.34 & 0.66 & $f_{gas}$ \\
 8615-12702 & Lin+2020 & 10.20 & -10.18 & -8.45 & -1.7 & \nodata & \nodata & \nodata \\
 8616-6104 & Lin+2020 & 10.77 & -10.52 & -9.09 & -1.6 & 0.47 & 0.53 & mixed \\
 8616-9102 & Lin+2020 & 10.44 & -9.79 & -8.74 & -1.0 & \nodata & \nodata & \nodata \\
 8616-12702 & Lin+2020 & 10.76 & -10.99 & -8.56 & -2.6 & 0.34 & 0.66 & $f_{gas}$ \\
 8618-9102 & Lin+2020 & 10.45 & -10.03 & -8.81 & -1.3 & \nodata & \nodata & \nodata \\
 8623-6104 & Lin+2020 & 11.28 & -10.11 & -8.83 & -1.4 & \nodata & \nodata & \nodata \\
 8623-12702 & Lin+2020 & 10.71 & -11.06 & -9.32 & -1.5 & 0.81 & 0.19 & SFE \\
 8655-3701 & Lin+2020 & 11.16 & -10.04 & -9.26 & -0.9 & \nodata & \nodata & \nodata \\
 8655-9102 & Lin+2020 & 10.31 & -10.03 & -8.54 & -1.6 & \nodata & \nodata & \nodata \\
 8655-12705 & Lin+2020 & 10.34 & -11.05 & -9.44 & -1.8 & 0.77 & 0.23 & SFE \\
 8728-3701 & Lin+2020 & 10.64 & -11.01 & -8.99 & -2.2 & 0.25 & 0.75 & $f_{gas}$ \\
 8950-12705 & Lin+2020 & 10.53 & -10.94 & -9.75 & -1.7 & 0.85 & 0.15 & SFE \\
 8952-6104 & Lin+2020 & 10.33 & -9.88 & -8.64 & -1.5 & \nodata & \nodata & \nodata \\
 8952-12701 & Lin+2020 & 10.17 & -10.55 & -9.19 & -1.9 & 0.49 & 0.51 & mixed \\
 8078-12703 & Thorp+2022 & 10.81 & -10.68 & -8.89 & -1.9 & 0.27 & 0.73 & $f_{gas}$ \\
 8082-9102 & Thorp+2022 & 10.74 & -10.02 & -9.04 & -1.1 & \nodata & \nodata & \nodata \\
 8083-12703 & Thorp+2022 & 10.46 & -10.14 & -8.65 & -1.5 & \nodata & \nodata & \nodata \\
 8085-12701 & Thorp+2022 & 10.43 & -9.85 & -8.07 & -1.6 & \nodata & \nodata & \nodata \\
 8085-3704 & Thorp+2022 & 10.72 & -10.07 & -9.02 & -1.1 & \nodata & \nodata & \nodata \\
 8616-9101 & Thorp+2022 & 11.10 & -10.35 & -8.40 & -1.6 & \nodata & \nodata & \nodata \\
 7977-9102 & Thorp+2022 & 10.93 & -10.10 & -8.91 & -1.3 & \nodata & \nodata & \nodata \\
 8078-6104 & Thorp+2022 & 10.41 & -10.18 & -8.87 & -1.3 & \nodata & \nodata & \nodata \\
 8241-12705 & Thorp+2022 & 10.40 & -10.39 & -8.99 & -1.4 & \nodata & \nodata & \nodata \\
 9195-9101 & Thorp+2022 & 10.74 & -10.35 & -8.91 & -1.4 & \nodata & \nodata & \nodata \\
 7975-6104 & Thorp+2022 & 11.01 & -10.40 & -9.25 & -1.3 & \nodata & \nodata & \nodata \\
 8623-1902 & Thorp+2022 & 10.17 & -11.17 & -9.25 & -2.1 & 0.09 & 0.91 & $f_{gas}$ \\
 9194-3702 & Thorp+2022 & 11.06 & -10.09 & -9.26 & -1.0 & \nodata & \nodata & \nodata \\
 9195-3702 & Thorp+2022 & 11.14 & -10.04 & -8.72 & -1.5 & \nodata & \nodata & \nodata \\
 9195-3703 & Thorp+2022 & 10.39 & -9.92 & -8.83 & -1.3 & \nodata & \nodata & \nodata \\
 9512-3704 & Thorp+2022 & 10.70 & -10.55 & -8.74 & -2.0 & 0.08 & 0.92 & $f_{gas}$ \\
 7964-1902 & Otter+2022 & 9.74 & -11.53 & -9.71 & -2.0 & 0.82 & 0.18 & SFE \\
 8080-3704 & Otter+2022 & 10.16 & -10.10 & -8.77 & -1.4 & \nodata & \nodata & \nodata \\
 8085-6104 & Otter+2022 & 9.98 & -10.32 & -8.62 & -1.7 & \nodata & \nodata & \nodata \\
 8086-3704 & Otter+2022 & 10.09 & -10.22 & -9.13 & -1.2 & \nodata & \nodata & \nodata \\
 8655-1902 & Otter+2022 & 9.57 & -11.30 & -9.95 & -1.5 & 1.00 & 0.00 & SFE \\
 8941-3701 & Otter+2022 & 10.22 & -11.86 & -10.32 & -2.0 & 0.98 & 0.02 & SFE \\
 8982-6104 & Otter+2022 & 10.67 & -9.98 & -9.03 & -1.1 & \nodata & \nodata & \nodata \\
 9088-9102 & Otter+2022 & 11.26 & -10.39 & -9.42 & -1.0 & \nodata & \nodata & \nodata \\
 9494-3701 & Otter+2022 & 10.02 & -11.27 & -8.86 & -2.5 & 0.00 & 1.00 & $f_{gas}$
\enddata
\tablecomments{$^{(a)}$ Sample categories based on observing programs. $^{(b)}$ The fraction of spaxels with suppressed sSFR driven by SFE depletion.  $^{(c)}$ The fraction of spaxels with suppressed sSFR driven by \fgas~ depletion. $^{(d)}$ Gas quenching mode. Only GV galaxies, i.e., those with log$_{10}$(sSFR/yr$^{-1}$) below -10.5, have the gas quenching mode classified.   }
\end{deluxetable*}

\clearpage
\startlongtable
\begin{deluxetable*}{llcccc}
	\tabletypesize{\scriptsize}
	\tablewidth{0pt}
	\tablecaption{Best-Fit Parameters for the Star formation History} \label{tab:SFH}
	\tablehead{
		\colhead{Plateifu} &
		\colhead{log$_{10}$($\frac{\rm sSFR}{\rm yr^{-1}}$)} &
		\colhead{$\alpha$} &
		\colhead{$\beta$} &
		\colhead{$\tau$ (Gyr)} &
		\colhead{$\tau_{decay}$ (Gyr)}
	}	
\startdata
7815-12705 & -10.42 & $0.10^{+<0.01}_{-<0.01}$ & $1.56^{+0.03}_{-0.03}$ & $12.98^{+0.02}_{-0.04}$ & \nodata \\
 7977-3703 & -9.95 & $0.10^{+<0.01}_{-<0.01}$ & $1.92^{+0.04}_{-0.03}$ & $12.98^{+0.02}_{-0.03}$ & \nodata \\
 7977-3704 & -10.75 & $14.57^{+0.13}_{-0.14}$ & $557.42^{+270.15}_{-230.76}$ & $10.19^{+0.02}_{-0.02}$ & $0.75^{+0.02}_{-0.01}$ \\
 7977-9101 & -11.10 & $1.04^{+0.01}_{-<0.01}$ & $2.31^{+10.98}_{-0.79}$ & $0.10^{+0.01}_{-<0.01}$ & $0.35^{+0.03}_{-0.11}$ \\
 7977-12705 & -10.39 & $10.70^{+0.11}_{-0.10}$ & $462.69^{+320.62}_{-223.83}$ & $9.76^{+0.03}_{-0.02}$ & $0.99^{+0.02}_{-0.02}$ \\
 8077-6104 & -10.08 & $1.51^{+0.23}_{-0.20}$ & $154.75^{+383.62}_{-120.74}$ & $6.34^{+0.36}_{-0.38}$ & $6.39^{+1.10}_{-0.85}$ \\
 8077-9101 & -10.50 & $0.77^{+0.04}_{-0.03}$ & $15.99^{+235.59}_{-15.42}$ & $0.16^{+0.11}_{-0.04}$ & $0.67^{+0.57}_{-0.27}$ \\
 8078-6103 & -10.14 & $2.25^{+0.05}_{-0.04}$ & $335.32^{+379.91}_{-199.49}$ & $7.43^{+0.06}_{-0.06}$ & $4.27^{+0.13}_{-0.10}$ \\
 8078-12701 & -10.54 & $9.20^{+0.08}_{-0.08}$ & $348.75^{+355.85}_{-197.11}$ & $9.14^{+0.02}_{-0.02}$ & $1.09^{+0.05}_{-0.02}$ \\
 8081-3704 & -9.59 & $0.10^{+<0.01}_{-<0.01}$ & $2.16^{+0.03}_{-0.02}$ & $12.73^{+0.01}_{-0.02}$ & \nodata \\
 8081-6102 & -11.19 & $11.91^{+0.26}_{-0.16}$ & $437.90^{+345.12}_{-203.63}$ & $8.55^{+0.07}_{-0.04}$ & $0.78^{+0.02}_{-0.01}$ \\
 8081-9101 & -10.28 & $1.94^{+0.11}_{-0.10}$ & $132.13^{+375.74}_{-99.66}$ & $5.29^{+0.24}_{-0.23}$ & $3.74^{+0.45}_{-0.16}$ \\
 8081-9102 & -10.42 & $2.30^{+0.08}_{-0.08}$ & $132.88^{+336.81}_{-99.68}$ & $4.10^{+0.17}_{-0.19}$ & $2.34^{+0.27}_{-0.08}$ \\
 8081-12703 & -11.13 & $16.02^{+0.30}_{-0.26}$ & $431.60^{+341.48}_{-210.25}$ & $10.12^{+0.04}_{-0.04}$ & $0.68^{+0.03}_{-0.02}$ \\
 8082-6103 & -9.92 & $0.10^{+<0.01}_{-<0.01}$ & $3.35^{+0.02}_{-0.02}$ & $13.00^{+<0.01}_{-0.01}$ & \nodata \\
 8082-12701 & -10.36 & $5.29^{+0.04}_{-0.04}$ & $665.24^{+235.80}_{-231.76}$ & $8.63^{+0.02}_{-0.02}$ & $1.83^{+0.02}_{-0.01}$ \\
 8082-12704 & -10.65 & $1.20^{+0.08}_{-0.04}$ & $0.68^{+30.99}_{-0.51}$ & $0.19^{+0.15}_{-0.06}$ & $0.44^{+0.35}_{-0.17}$ \\
 8083-6101 & -11.18 & $17.53^{+0.43}_{-0.33}$ & $580.39^{+250.13}_{-245.66}$ & $10.37^{+0.05}_{-0.04}$ & $0.63^{+0.01}_{-0.01}$ \\
 8083-9101 & -10.54 & $3.06^{+0.08}_{-0.08}$ & $224.78^{+414.20}_{-145.93}$ & $6.21^{+0.11}_{-0.11}$ & $2.49^{+0.12}_{-0.07}$ \\
 8083-12702 & -10.56 & $0.10^{+<0.01}_{-<0.01}$ & $0.48^{+0.01}_{-0.01}$ & $12.86^{+0.10}_{-0.19}$ & \nodata \\
 8084-3702 & -9.80 & $0.11^{+0.02}_{-0.01}$ & $332.73^{+10.21}_{-10.08}$ & $13.00^{+<0.01}_{-<0.01}$ & \nodata \\
 8084-6103 & -10.65 & $12.84^{+0.28}_{-0.30}$ & $302.56^{+395.85}_{-169.26}$ & $9.76^{+0.05}_{-0.06}$ & $0.83^{+0.05}_{-0.03}$ \\
 8084-12705 & -10.48 & $3.16^{+0.12}_{-0.12}$ & $165.61^{+387.35}_{-116.37}$ & $6.86^{+0.14}_{-0.16}$ & $2.68^{+0.23}_{-0.11}$ \\
 8086-9101 & -10.83 & $5.21^{+0.12}_{-0.11}$ & $216.39^{+386.55}_{-145.59}$ & $7.08^{+0.10}_{-0.08}$ & $1.56^{+0.11}_{-0.04}$ \\
 8155-6101 & -11.35 & $7.30^{+0.19}_{-0.14}$ & $250.38^{+342.34}_{-169.16}$ & $5.91^{+0.12}_{-0.10}$ & $0.91^{+0.06}_{-0.02}$ \\
 8155-6102 & -10.00 & $0.42^{+0.09}_{-0.10}$ & $131.49^{+381.44}_{-103.17}$ & $5.20^{+0.35}_{-0.38}$ & \nodata \\
 8156-3701 & -9.65 & $0.11^{+0.01}_{-<0.01}$ & $1.41^{+0.02}_{-0.02}$ & $12.65^{+0.08}_{-0.17}$ & \nodata \\
 8241-3703 & -9.86 & $0.11^{+0.02}_{-0.01}$ & $1.55^{+0.04}_{-0.03}$ & $12.79^{+0.15}_{-0.31}$ & \nodata \\
 8241-3704 & -9.79 & $0.10^{+0.01}_{-<0.01}$ & $0.53^{+0.01}_{-0.01}$ & $12.30^{+0.20}_{-0.49}$ & \nodata \\
 8450-6102 & -9.79 & $0.15^{+0.11}_{-0.04}$ & $1.87^{+0.12}_{-0.07}$ & $12.06^{+0.61}_{-0.96}$ & \nodata \\
 8615-3703 & -9.79 & $0.11^{+0.01}_{-0.01}$ & $192.17^{+5.08}_{-4.09}$ & $13.00^{+<0.01}_{-<0.01}$ & \nodata \\
 8615-9101 & -10.55 & $0.93^{+0.08}_{-0.07}$ & $9.63^{+196.82}_{-9.21}$ & $0.43^{+0.25}_{-0.20}$ & $1.18^{+1.00}_{-0.49}$ \\
 8615-12702 & -10.18 & $0.10^{+0.01}_{-<0.01}$ & $0.90^{+0.03}_{-0.03}$ & $12.83^{+0.13}_{-0.24}$ & \nodata \\
 8616-6104 & -10.52 & $1.43^{+0.23}_{-0.20}$ & $54.25^{+364.68}_{-49.69}$ & $2.90^{+0.91}_{-0.72}$ & $3.20^{+1.64}_{-0.28}$ \\
 8616-9102 & -9.79 & $0.11^{+0.01}_{-0.01}$ & $1.85^{+0.04}_{-0.04}$ & $12.92^{+0.06}_{-0.12}$ & \nodata \\
 8616-12702 & -10.99 & $11.43^{+0.20}_{-0.16}$ & $360.75^{+395.68}_{-184.19}$ & $8.80^{+0.06}_{-0.04}$ & $0.84^{+0.03}_{-0.02}$ \\
 8618-9102 & -10.03 & $0.12^{+0.03}_{-0.02}$ & $0.76^{+0.06}_{-0.06}$ & $11.72^{+0.86}_{-1.37}$ & \nodata \\
 8623-6104 & -10.11 & $0.30^{+0.01}_{-0.02}$ & $94.91^{+373.45}_{-84.90}$ & $0.11^{+0.02}_{-<0.01}$ & $3.67^{+1.97}_{-0.78}$ \\
 8623-12702 & -11.06 & $0.58^{+0.01}_{-0.01}$ & $92.07^{+380.85}_{-73.93}$ & $0.12^{+0.01}_{-0.01}$ & $0.62^{+0.09}_{-0.08}$ \\
 8655-3701 & -10.04 & $0.21^{+0.10}_{-0.05}$ & $56.33^{+315.23}_{-50.88}$ & $1.38^{+0.79}_{-0.55}$ & \nodata \\
 8655-9102 & -10.03 & $0.15^{+0.09}_{-0.04}$ & $0.52^{+0.13}_{-0.08}$ & $9.30^{+2.66}_{-3.41}$ & \nodata \\
 8655-12705 & -11.05 & $18.54^{+0.37}_{-0.45}$ & $489.81^{+328.48}_{-235.54}$ & $10.03^{+0.03}_{-0.05}$ & $0.58^{+0.02}_{-0.01}$ \\
 8728-3701 & -11.01 & $13.53^{+0.23}_{-0.18}$ & $565.89^{+264.50}_{-269.62}$ & $9.59^{+0.04}_{-0.03}$ & $0.76^{+0.01}_{-0.01}$ \\
 8950-12705 & -10.94 & $13.51^{+0.11}_{-0.11}$ & $676.38^{+204.00}_{-214.39}$ & $10.27^{+0.01}_{-0.02}$ & $0.81^{+0.01}_{-0.01}$ \\
 8952-6104 & -9.88 & $0.11^{+0.02}_{-0.01}$ & $3.09^{+0.05}_{-0.05}$ & $12.96^{+0.03}_{-0.07}$ & \nodata \\
 8952-12701 & -10.55 & $24.85^{+0.17}_{-0.18}$ & $440.71^{+294.26}_{-196.89}$ & $10.69^{+0.01}_{-0.01}$ & $0.47^{+0.02}_{-0.01}$ \\
 8078-12703 & -10.68 & $5.17^{+0.12}_{-0.18}$ & $290.25^{+377.49}_{-180.13}$ & $7.72^{+0.08}_{-0.12}$ & $1.72^{+0.06}_{-0.05}$ \\
 8082-9102 & -10.02 & $0.13^{+0.01}_{-0.01}$ & $58.45^{+366.45}_{-52.91}$ & $0.13^{+0.03}_{-0.02}$ & \nodata \\
 8083-12703 & -10.14 & $0.10^{+0.01}_{-<0.01}$ & $1.21^{+0.03}_{-0.03}$ & $12.87^{+0.10}_{-0.19}$ & \nodata \\
 8085-12701 & -9.85 & $0.10^{+<0.01}_{-<0.01}$ & $2.04^{+0.02}_{-0.02}$ & $12.99^{+0.01}_{-0.02}$ & \nodata \\
 8085-3704 & -10.07 & $0.27^{+0.01}_{-0.01}$ & $107.16^{+409.45}_{-96.67}$ & $0.12^{+0.02}_{-0.01}$ & $5.54^{+1.39}_{-0.88}$ \\
 8616-9101 & -10.35 & $0.73^{+0.02}_{-0.01}$ & $7.34^{+103.85}_{-7.00}$ & $0.13^{+0.03}_{-0.02}$ & $0.60^{+0.35}_{-0.19}$ \\
 7977-9102 & -10.10 & $0.19^{+0.09}_{-0.06}$ & $0.15^{+0.09}_{-0.04}$ & $5.70^{+4.91}_{-3.74}$ & \nodata \\
 8078-6104 & -10.18 & $0.15^{+0.04}_{-0.03}$ & $0.11^{+0.04}_{-0.01}$ & $1.01^{+3.42}_{-0.80}$ & \nodata \\
 8241-12705 & -10.39 & $4.88^{+0.17}_{-0.16}$ & $289.46^{+345.13}_{-189.45}$ & $9.11^{+0.08}_{-0.08}$ & $2.16^{+0.11}_{-0.08}$ \\
 9195-9101 & -10.35 & $0.44^{+0.02}_{-0.01}$ & $48.07^{+339.32}_{-44.17}$ & $0.12^{+0.05}_{-0.01}$ & $1.37^{+1.19}_{-0.32}$ \\
 7975-6104 & -10.40 & $8.01^{+0.13}_{-0.13}$ & $472.93^{+355.70}_{-252.44}$ & $8.86^{+0.03}_{-0.05}$ & $1.21^{+0.03}_{-0.02}$ \\
 8623-1902 & -11.17 & $17.11^{+0.38}_{-0.39}$ & $327.86^{+365.90}_{-187.85}$ & $9.96^{+0.05}_{-0.05}$ & $0.64^{+0.05}_{-0.02}$ \\
 9194-3702 & -10.09 & $18.87^{+0.45}_{-0.45}$ & $26.28^{+1.45}_{-1.06}$ & $11.34^{+0.02}_{-0.03}$ & $0.96^{+0.02}_{-0.02}$ \\
 9195-3702 & -10.04 & $10.10^{+0.17}_{-0.18}$ & $321.82^{+409.70}_{-180.74}$ & $9.28^{+0.04}_{-0.05}$ & $1.01^{+0.04}_{-0.03}$ \\
 9195-3703 & -9.92 & $0.10^{+<0.01}_{-<0.01}$ & $1.34^{+0.01}_{-0.01}$ & $12.98^{+0.02}_{-0.04}$ & \nodata \\
 9512-3704 & -10.55 & $5.39^{+0.23}_{-0.21}$ & $215.90^{+413.64}_{-153.88}$ & $7.27^{+0.15}_{-0.14}$ & $1.56^{+0.12}_{-0.06}$ \\
 7964-1902 & -11.53 & $35.08^{+0.77}_{-0.75}$ & $59.54^{+5.34}_{-4.77}$ & $11.25^{+0.03}_{-0.03}$ & $0.48^{+0.01}_{-0.01}$ \\
 8080-3704 & -10.10 & $16.14^{+0.05}_{-0.08}$ & $355.51^{+150.73}_{-62.86}$ & $11.02^{+0.01}_{-0.01}$ & $0.75^{+0.01}_{-0.01}$ \\
 8085-6104 & -10.32 & $3.34^{+0.20}_{-0.15}$ & $505.39^{+312.16}_{-258.09}$ & $10.41^{+0.05}_{-0.04}$ & $3.71^{+0.19}_{-0.22}$ \\
 8086-3704 & -10.22 & $2.88^{+0.43}_{-0.36}$ & $156.28^{+444.18}_{-114.10}$ & $7.64^{+0.42}_{-0.40}$ & $3.39^{+0.45}_{-0.39}$ \\
 8655-1902 & -11.30 & $59.67^{+0.62}_{-1.11}$ & $6.72^{+0.21}_{-0.36}$ & $12.34^{+0.01}_{-0.02}$ & $0.62^{+0.01}_{-0.01}$ \\
 8941-3701 & -11.86 & $42.93^{+0.41}_{-0.39}$ & $567.93^{+267.51}_{-250.04}$ & $11.27^{+0.02}_{-0.02}$ & $0.29^{+0.01}_{-0.01}$ \\
 8982-6104 & -9.98 & $5.81^{+0.18}_{-0.19}$ & $285.19^{+387.14}_{-196.89}$ & $9.21^{+0.09}_{-0.09}$ & $1.81^{+0.11}_{-0.07}$ \\
 9088-9102 & -10.39 & $11.24^{+0.18}_{-0.18}$ & $313.43^{+349.41}_{-182.05}$ & $9.47^{+0.04}_{-0.04}$ & $0.92^{+0.05}_{-0.03}$ \\
 9494-3701 & -11.27 & $70.34^{+0.63}_{-0.57}$ & $20.20^{+0.99}_{-0.96}$ & $12.43^{+0.01}_{-0.01}$ & $0.41^{+<0.01}_{-<0.01}$
\enddata
\tablecomments{$\alpha$, $\beta$, and $\tau$ describe the parameters in the SFH model given in Eq. \ref{eq:sfh}. Galaxies whose SFR does not drop below 1/e of its peak value within 10 Gyr from the present day are omitted.}
\end{deluxetable*}

\appendix
\section{Star Formation History}
The methods modeling the star formation histories are described in \S2.3. Figure \ref{fig:SFH00} shows the spectral fit and the corresponding SFH reconstruction for all 71 galaxies used in this work.
\renewcommand{\thefigure}{A\arabic{figure}}

\begin{figure*}
\figurenum{A1}
\plotone{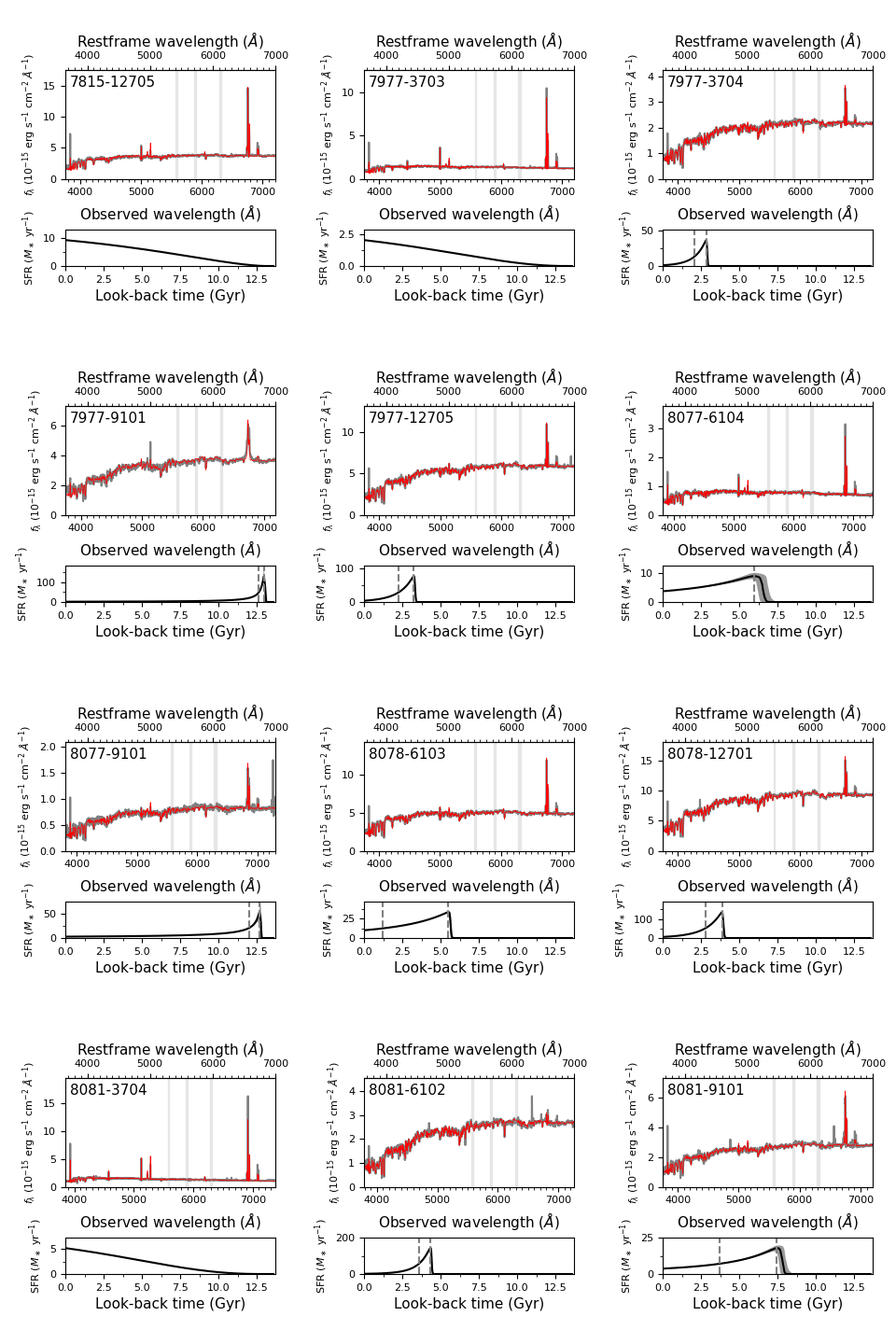}
\caption{Same as Figure \ref{fig:sfh} but shown for all 71 galaxies used in this work. Objects are displayed (from left to right, top to bottom) in the same order as in the Table \ref{tab:sample}. The complete figure set (6 images) is available in the online journal.\label{fig:SFH00}}
\end{figure*}

\figsetgrpstart
\figsetgrpnum{A1.2}
\figsetgrptitle{SFH of 71 galaxies}
\figsetplot{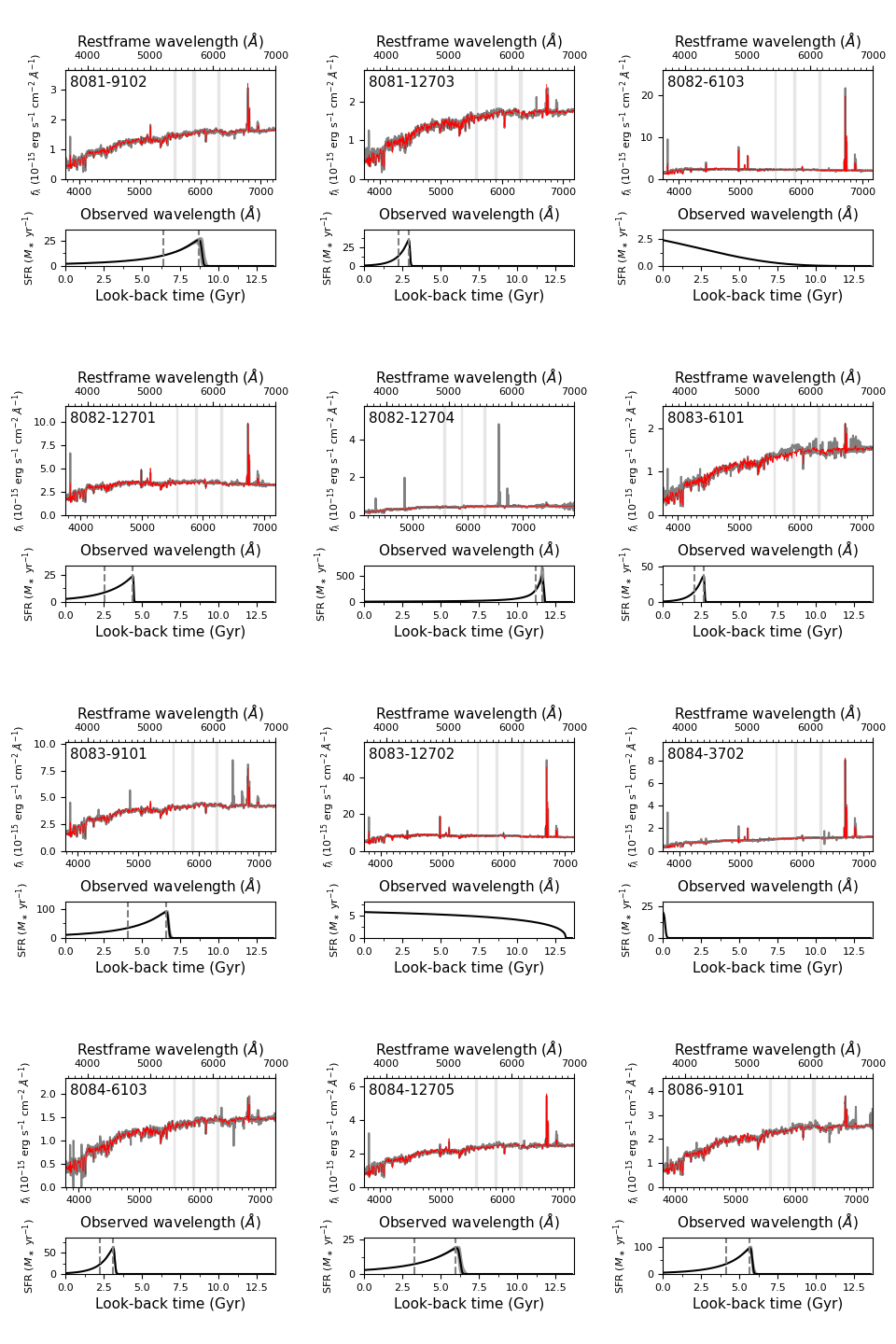}
\figsetgrpnote{Continued.}
\figsetgrpend

\figsetgrpstart
\figsetgrpnum{A1.3}
\figsetgrptitle{SFH of 71 galaxies}
\figsetplot{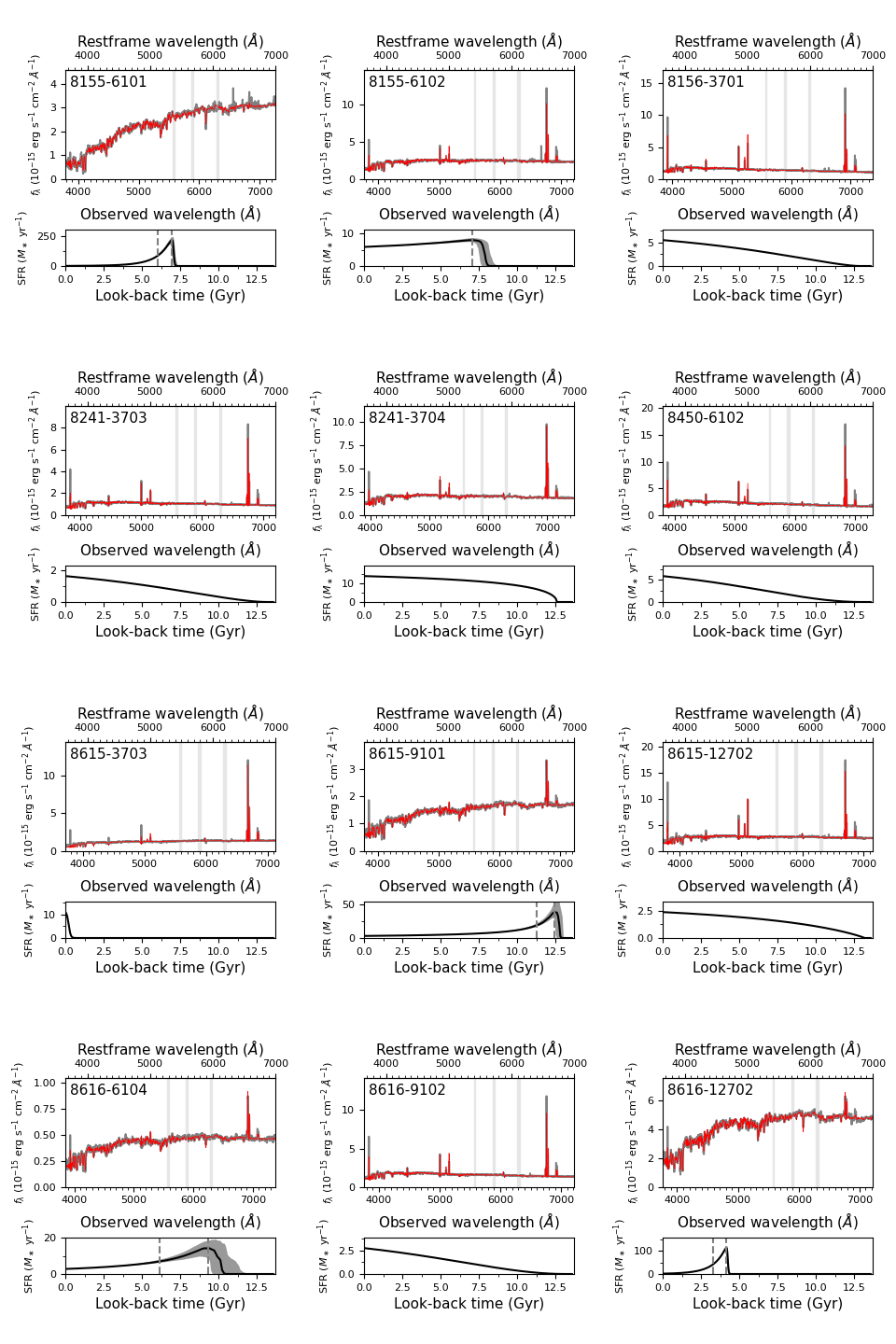}
\figsetgrpnote{Continued.}
\figsetgrpend

\figsetgrpstart
\figsetgrpnum{A1.4}
\figsetgrptitle{SFH of 71 galaxies}
\figsetplot{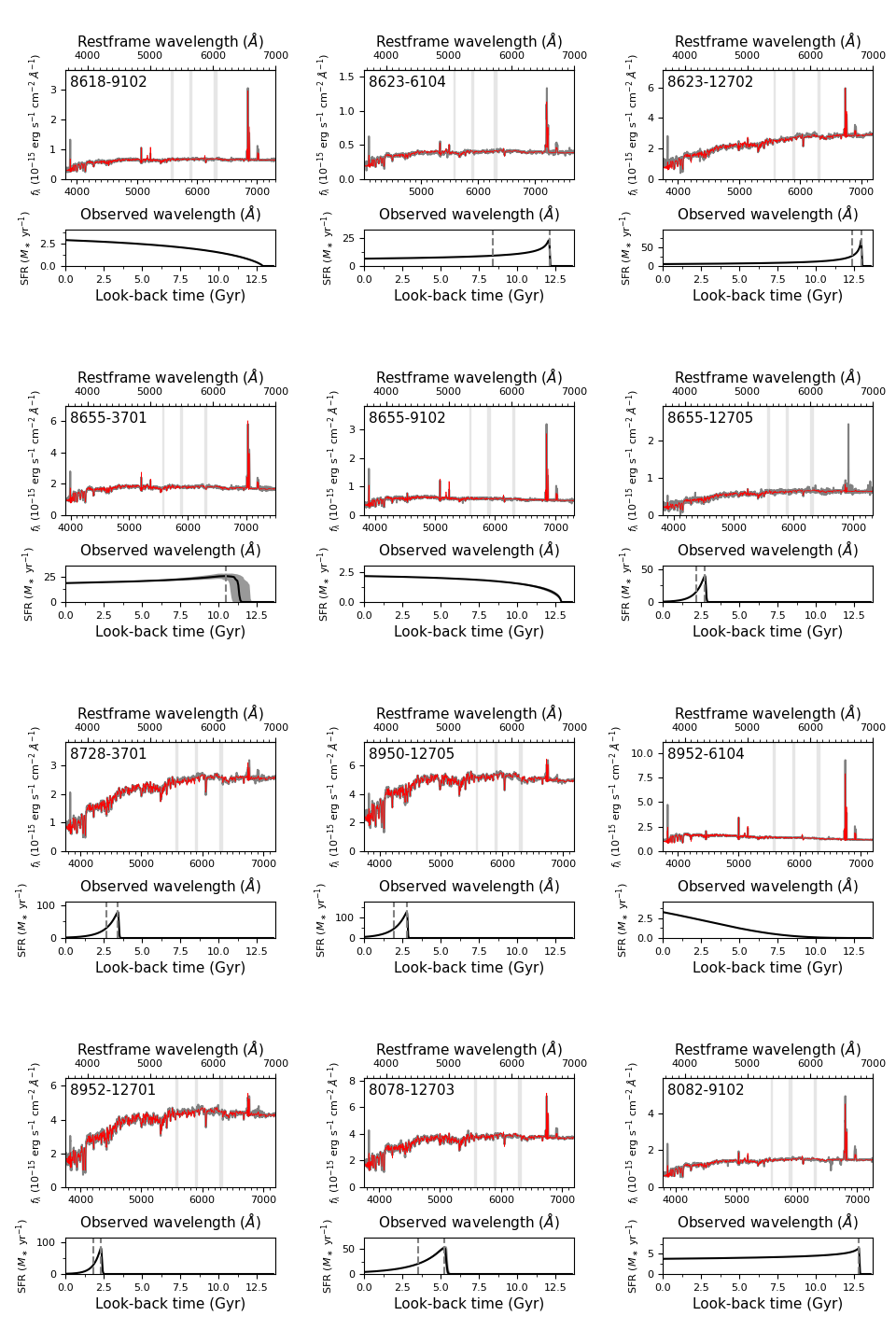}
\figsetgrpnote{Continued.}
\figsetgrpend

 \figsetgrpstart
\figsetgrpnum{A1.5}
\figsetgrptitle{SFH of 71 galaxies}
\figsetplot{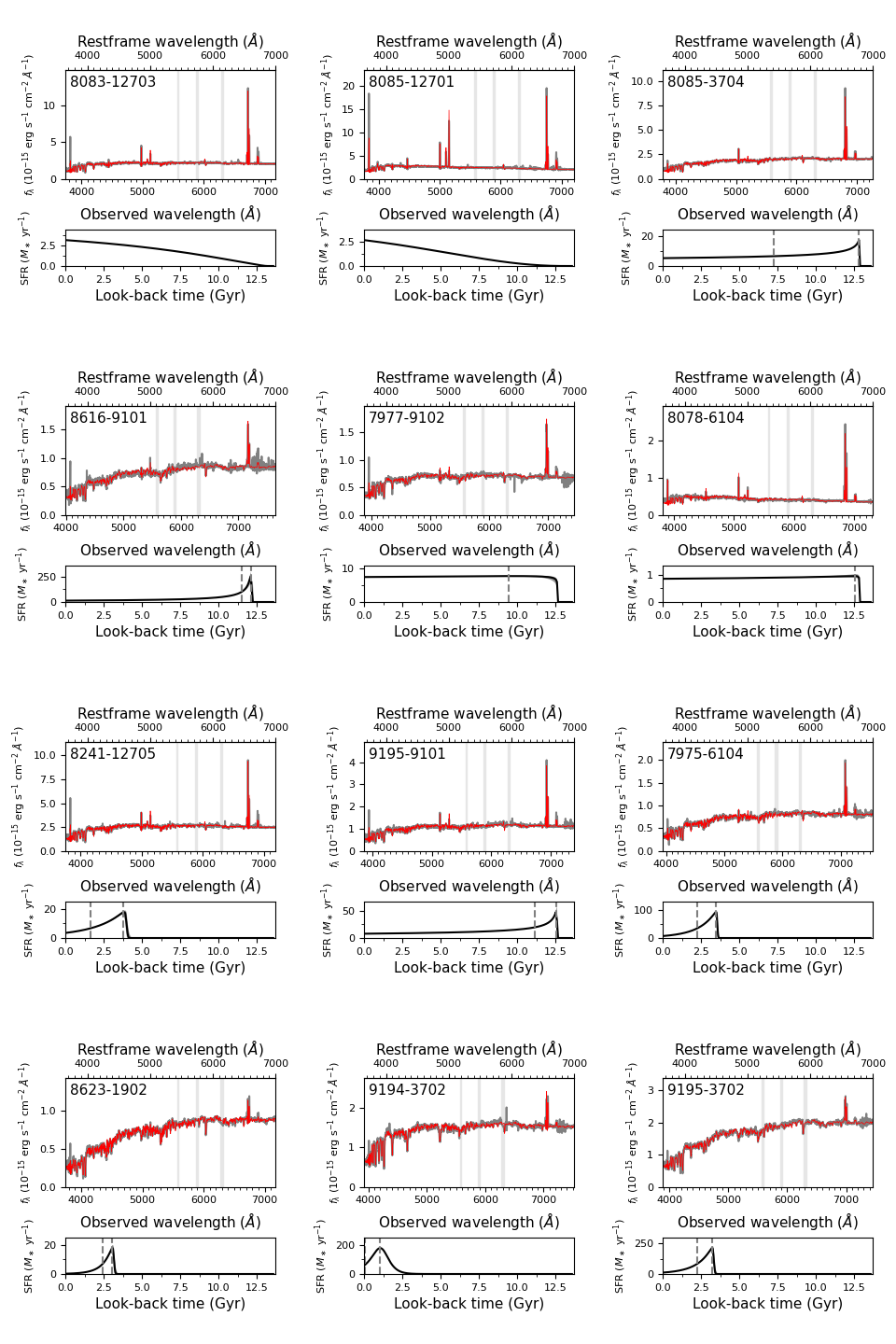}
\figsetgrpnote{Continued.}
\figsetgrpend

\figsetgrpstart
\figsetgrpnum{A1.6}
\figsetgrptitle{SFH of 71 galaxies}
\figsetplot{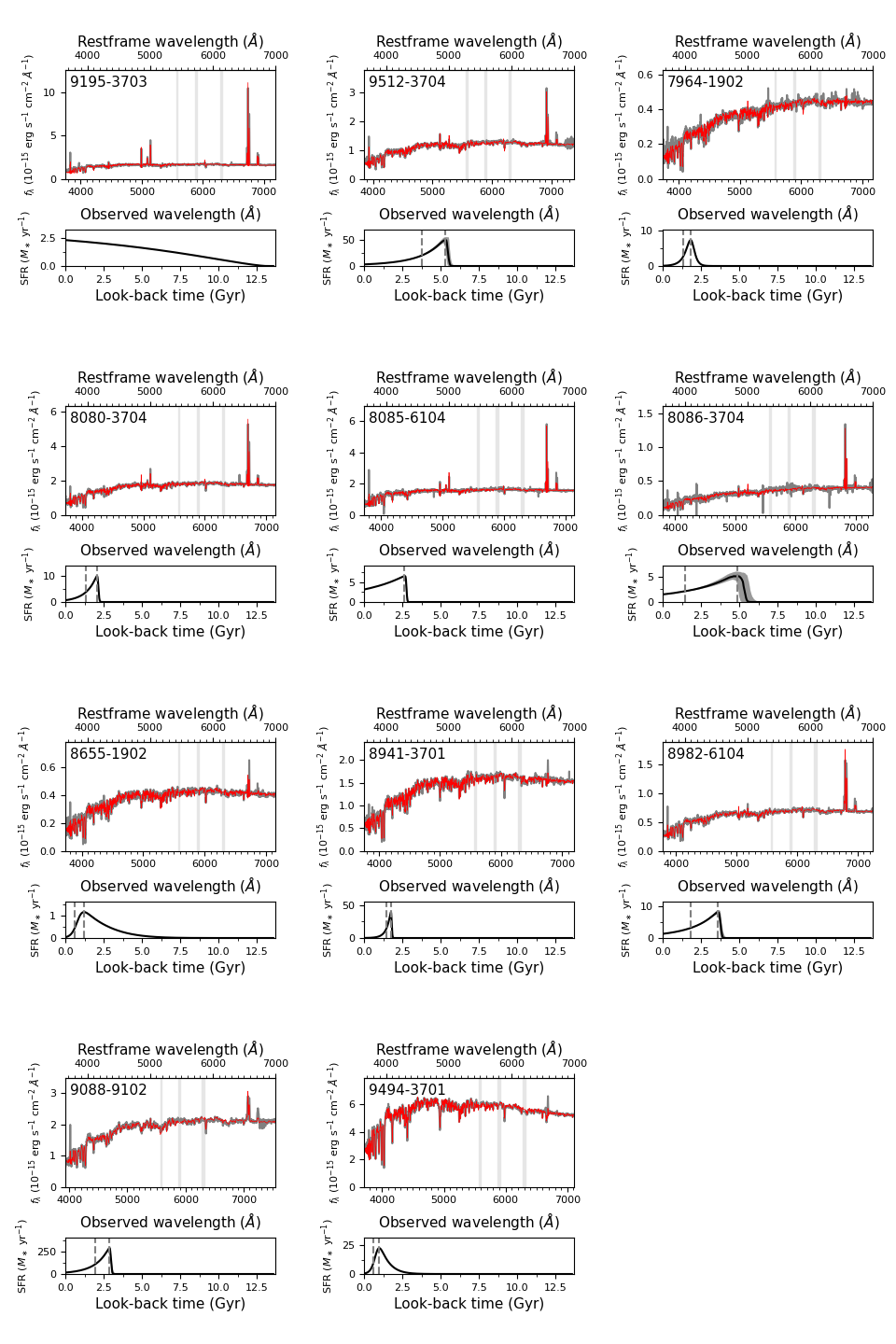}
\figsetgrpnote{Continued.}
\figsetgrpend


\begin{thebibliography}{}
\bibitem[Aguado et al.(2019)]{aqu19} Aguado, D.~S., Ahumada, R., Almeida, A., et al.\ 2019, \apjs, 240, 23
\bibitem[Baldwin, Phillips, \& Terlevich (1981)]{bal81} Baldwin, J.~A., Phillips, M.~M., \& Terlevich, R.\ 1981, \pasp, 93, 5 
\bibitem[Baker et al.(2022)]{bak22} Baker, W.~M., Maiolino, R., Bluck, A.~F.~L., et al.\ 2022, \mnras, 510, 3622
\bibitem[Barrera-Ballesteros et al.(2020)]{bar20} Barrera-Ballesteros, J.~K., Utomo, D., Bolatto, A.~D., et al.\ 2020, \mnras, 492, 2651
\bibitem[Barrera-Ballesteros et al.(2021)]{bar21} Barrera-Ballesteros, J.~K., Heckman, T., S{\'a}nchez, S.~F., et al.\ 2021, \apj, 909, 131
\bibitem[Baxter et al.(2025)]{bax25} Baxter, D.~C., Fillingham, S.~P., Coil, A.~L., et al.\ 2025, \apj, 979, 41. 
\bibitem[Bazzi et al.(2025)]{baz25} Bazzi, Z., Colombo, D., Bigiel, F., et al.\ 2025, \aap, 697, A149. 
\bibitem[Behroozi et al.(2013)]{beh13} Behroozi, P.~S., Wechsler, R.~H., \& Conroy, C.\ 2013, \apj, 770, 1, 57
\bibitem[Belfiore et al.(2016)]{bel16} Belfiore, F., Maiolino, R., Maraston, C., et al.\ 2016, \mnras, 461, 3111 
\bibitem[Belfiore et al.(2017)]{bel17} Belfiore, F., Maiolino, R., Maraston, C., et al.\ 2017, \mnras, 466, 2570
\bibitem[Belfiore et al.(2018)]{bel18} Belfiore, F., Maiolino, R., Bundy, K., et al.\ 2018, \mnras, 477, 3014 
\bibitem[Belfiore et al.(2016)]{bel16} Belfiore, F., Maiolino, R., Maraston, C., et al.\ 2016, \mnras, 461, 3111 
\bibitem[Belfiore et al.(2019)]{bel19} Belfiore, F., Westfall, K.~B., Schaefer, A., et al.\ 2019, \aj, 158, 4, 160. 
\bibitem[Bolatto et al.(2013)]{bol13} Bolatto, A.~D., Wolfire, M., \& Leroy, A.~K.\ 2013, \araa, 51, 207 
\bibitem[Bravo et al.(2023)]{bra23} Bravo, M., Robotham, A.~S.~G., Lagos, C. del P., et al.\ 2023, \mnras
\bibitem[Bremer et al.(2018)]{bre18} Bremer, M.~N., Phillipps, S., Kelvin, L.~S., et al.\ 2018, \mnras
\bibitem[Brinchmann et al.(2004)]{bri04} Brinchmann, J., Charlot, S., White, S.~D.~M., et al.\ 2004, \mnras, 351, 1151
\bibitem[Brownson et al.(2020)]{bro20} Brownson, S., Belfiore, F., Maiolino, R., et al.\ 2020, \mnras, 498, L66.
\bibitem[Bruzual \& Charlot(2003)]{bru03} Bruzual, G. \& Charlot, S.\ 2003, \mnras, 344, 1000. 
\bibitem[Bundy et al.(2015)]{bun15} Bundy, K., Bershady, M. A., Law, D. R., et al. 2015, \apj, 798, 7
\bibitem[Cano-D{\'{\i}}az et al.(2019)]{can19} Cano-D{\'{\i}}az, M., {\'A}vila-Reese, V., S{\'a}nchez, S.~F., et al.\ 2019, \mnras, 1830
\bibitem[Caplar \& Tacchella(2019)]{cap19} Caplar, N. \& Tacchella, S.\ 2019, \mnras, 487, 3845. 
\bibitem[Cardelli et al.(1989)]{car89} Cardelli, J.~A., Clayton, G.~C., \& Mathis, J.~S.\ 1989, \apj, 345, 245
\bibitem[Carnall et al.(2018)]{car18} Carnall, A.~C., McLure, R.~J., Dunlop, J.~S., et al.\ 2018, \mnras, 480, 4379.
\bibitem[Carnall et al.(2019)]{car19} Carnall, A.~C., McLure, R.~J., Dunlop, J.~S., et al.\ 2019, \mnras, 490, 417. 
\bibitem[Charlot \& Fall(2000)]{cha00} Charlot, S. \& Fall, S.~M.\ 2000, \apj, 539, 718. 
\bibitem[Chauke et al.(2019)]{cha19} Chauke, P., van der Wel, A., Pacifici, C., et al.\ 2019, \apj, 877, 48. 
\bibitem[Chen et al.(2015)]{che15} Chen, H., Gao, Y., Braine, J., et al.\ 2015, \apj, 810, 140
\bibitem[Chen et al.(2019)]{che19} Chen, Y.-M., Shi, Y., Wild, V., et al.\ 2019, \mnras, 489, 5709. 
\bibitem[Chevallard \& Charlot(2016)]{che16} Chevallard, J. \& Charlot, S.\ 2016, \mnras, 462, 1415. 
\bibitem[Chiang et al.(2021)]{chi21} Chiang, I.-D., Sandstrom, K.~M., Chastenet, J., et al.\ 2021, \apj, 907, 29. 
\bibitem[Chiang et al.(2025)]{chi25} Chiang, I.-D., Hirashita, H., Chastenet, J., et al.\ 2025, \mnras, 536, 3, 2392. 
\bibitem[Cid Fernandes et al.(2010)]{cid10} Cid Fernandes, R., Stasi{\'n}ska, G., Schlickmann, M.~S., et al.\ 2010, \mnras, 403, 1036 
\bibitem[Cid Fernandes et al.(2011)]{cid11} Cid Fernandes, R., Stasi{\'n}ska, G., Mateus, A., \& Vale Asari, N.\ 2011, \mnras, 413, 1687 
\bibitem[Coenda et al.(2018)]{coe18} Coenda, V., Mart{\'\i}nez, H.~J., \& Muriel, H.\ 2018, \mnras, 473, 5617. 
\bibitem[Colombo et al.(2020)]{col20} Colombo, D., Sanchez, S.~F., Bolatto, A.~D., et al.\ 2020, \aap, 644, A97. 
\bibitem[Colombo et al.(2025)]{col25} Colombo, D., Kalinova, V., Bazzi, Z., et al.\ 2025, \aap, 699, A367. doi:10.1051/0004-6361/202453217
\bibitem[Croton et al.(2006)]{cro06} Croton, D.~J., Springel, V., White, S.~D.~M., et al.\ 2006, \mnras, 365, 11 
\bibitem[Daddi et al.(2007)]{dad07} Daddi, E., Dickinson, M., Morrison, G., et al.\ 2007, \apj, 670, 156
\bibitem[Diemer et al.(2017)]{die17} Diemer, B., Sparre, M., Abramson, L.~E., et al.\ 2017, \apj, 839, 26. 
\bibitem[Di Matteo et al.(2005)]{di05} Di Matteo, T., Springel, V., \& Hernquist, L.\ 2005, \nat, 433, 7026, 604.
\bibitem[Ellison et al.(2021a)]{ell21a} Ellison, S.~L., Lin, L., Thorp, M.~D., et al.\ 2021a, \mnras, 501, 4777
\bibitem[Ellison et al.(2021b)]{ell21b} Ellison, S.~L., Lin, L., Thorp, M.~D., et al.\ 2021b, \mnras, 502, L6. 
\bibitem[Ellison et al.(2024)]{ell24} Ellison, S.~L., Pan, H.-A., Bluck, A.~F.~L., et al.\ 2024, \mnras, 527, 10201. 
\bibitem[Enia et al.(2020)]{eni20} Enia, A., Rodighiero, G., Morselli, L., et al.\ 2020, \mnras, 493, 4107.
\bibitem[Fabian(2012)]{fab12} Fabian, A.~C.\ 2012, \araa, 50, 455 
\bibitem[Falc{\'o}n-Barroso et al.(2011)]{fal11} Falc{\'o}n-Barroso, J., S{\'a}nchez-Bl{\'a}zquez, P., Vazdekis, A., et al.\ 2011, \aap, 532, A95. 
\bibitem[Federrath \& Klessen(2012)]{fed12} Federrath, C. \& Klessen, R.~S.\ 2012, \apj, 761, 156. 
bibitem[Falc{\'o}n-Barroso et al.(2011)]{fal11} Falc{\'o}n-Barroso, J., S{\'a}nchez-Bl{\'a}zquez, P., Vazdekis, A., et al.\ 2011, \aap, 532, A95. 
\bibitem[Feldmann et al.(2012)]{feld12} Feldmann, R., Gnedin, N.~Y., \& Kravtsov, A.~V.\ 2012, \apj, 747, 124.
\bibitem[Gallagher et al.(2018)]{gal18} Gallagher, M.~J., Leroy, A.~K., Bigiel, F., et al.\ 2018, \apj, 858, 90
\bibitem[Garc{\'\i}a-Burillo et al.(2012)]{gar12} Garc{\'\i}a-Burillo, S., Usero, A., Alonso-Herrero, A., et al.\ 2012, \aap, 539, A8.
\bibitem[Gladders et al.(2013)]{gla13} Gladders, M.~D., Oemler, A., Dressler, A., et al.\ 2013, \apj, 770, 1, 64.
\bibitem[Herrera-Camus et al.(2019)]{her19} Herrera-Camus, R., Tacconi, L., Genzel, R., et al.\ 2019, \apj, 871, 1, 37
\bibitem[Hamadouche et al.(2023)]{ham23} Hamadouche, M.~L., Carnall, A.~C., McLure, R.~J., et al.\ 2023, \mnras
\bibitem[Hsieh et al.(2017)]{hsi17}  Hsieh, B.~C., Lin, L., Lin, J.~H., et al.\ 2017, \apjl, 851, L24 
\bibitem[Ivezi{\'c} et al.(2014)]{ive14} Ivezi{\'c}, {\v{Z}}., Connolly, A.~J., VanderPlas, J.~T., et al.\ 2014, Statistics, Data Mining, and Machine Learning in Astronomy: A Practical Python Guide for the Analysis of Survey Data, by Ivezi{\'c}, {\v{Z}}eljko; Connolly, Andrew; Vanderplas, Jacob T.; Gray, Alexander, 2014. Princeton: Princeton University Press. OCLC: 979780267. ISBN: 9781400848911.. 
\bibitem[Iyer et al.(2019)]{iye19} Iyer, K.~G., Gawiser, E., Faber, S.~M., et al.\ 2019, \apj, 879, 116. 
\bibitem[Iyer et al.(2020)]{iye20} Iyer, K.~G., Tacchella, S., Genel, S., et al.\ 2020, \mnras, 498, 430. 
\bibitem[Iyer et al.(2024)]{iye24} Iyer, K.~G., Speagle, J.~S., Caplar, N., et al.\ 2024, \apj, 961, 53. 
\bibitem[Jian et al.(2020)]{jia20} Jiang, X.-J., Greve, T.~R., Gao, Y., et al.\ 2020, \mnras, 494, 1276. 
\bibitem[Jian et al.(2022)]{jia22} Jian, H.-Y., Lin, L., Hsieh, B.-C., et al.\ 2022, \apj, 926, 2, 115. 
\bibitem[Kaviraj et al.(2011)]{kav11} Kaviraj, S., Schawinski, K., Silk, J., et al.\ 2011, \mnras
\bibitem[Kauffmann et al.(2003)]{kau03} Kauffmann, G., Heckman, T.~M., Tremonti, C., et al.\ 2003, 
\bibitem[Kelkar et al.(2019)]{kel19} Kelkar, K., Gray, M.~E., Arag{\'o}n-Salamanca, A., et al.\ 2019, \mnras, 486, 1, 868.
\bibitem[Kennicutt(1998)]{ken98} Kennicutt, R.~C.\ 1998, \apj, 498, 2, 541. doi:10.1086/305588
\bibitem[Kewley et al.(2001)]{kew01} Kewley, L.~J., Dopita, M.~A., Sutherland, R. S., Heisler, C.~A., \& Trevena, J.\ 2001, \apj, 556, 121 
\bibitem[Lacerda et al.(2020)]{lac20} Lacerda, E.~A.~D., S{\'a}nchez, S.~F., Cid Fernandes, R., et al.\ 2020, \mnras, 492, 3, 3073. 
\bibitem[Larson et al.(1980)]{lar80} Larson, R. B., Tinsley, B. M., \& Caldwell, C. N. 1980, \apj, 237, 692
\bibitem[Law et al.(2016)]{law16} Law, D.~R., Cherinka, B., Yan, R., et al.\ 2016, AJ, 152, 83
\bibitem[Leroy et al.(2025)]{ler25} Leroy, A.~K., Sun, J., Meidt, S., et al.\ 2025, \apj, 985, 1, 14. 
\bibitem[Leung et al.(2024)]{leu24} Leung, H.-H., Wild, V., Papathomas, M., et al.\ 2024, \mnras, 528, 3, 4029.
\bibitem[Lin et al.(2017)]{lin17} Lin, L., Belfiore, F., Pan, H.-A., et al.\ 2017, \apj, 851, 18 
\bibitem[Lin et al.(2019b)]{lin19b} Lin, L., Pan, H.-A., Ellison, S.~L., et al.\ 2019b, \apjl, 884, L33
\bibitem[Lin et al.(2020)]{lin20} Lin, L., Ellison, S.~L., Pan, H.-A., et al.\ 2020, \apj, 903, 145. 
\bibitem[Lin et al.(2022)]{lin22} Lin, L., Ellison, S.~L., Pan, H.-A., et al.\ 2022, \apj, 926, 175
\bibitem[Lin et al.(2024)]{lin24} Lin, L., Pan, H.-A., Ellison, S.~L., et al.\ 2024, \apj, 963, 115. 
\bibitem[Martig et al.(2009)]{mar09} Martig, M., Bournaud, F., Teyssier, R., \& Dekel, A.\ 2009, \apj, 707, 250 
\bibitem[Martin et al.(2007)]{mar07} Martin, D.~C., Wyder, T.~K., Schiminovich, D., et al.\ 2007, \apjs, 173, 342 
\bibitem[Muzzin et al.(2014)]{muz14} Muzzin, A., van der Burg, R.~F.~J., McGee, S.~L., et al.\ 2014, \apj, 796, 65.
\bibitem[Nogueira-Cavalcante et al.(2019)]{nog19} Nogueira-Cavalcante, J.~P., Dupke, R., Coelho, P., et al.\ 2019, \aap, 630, A88. 
\bibitem[Morselli et al.(2020)]{mor20} Morselli, L., Rodighiero, G., Enia, A., et al.\ 2020, \mnras, 496, 4606
\bibitem[Narayanan et al.(2012)]{nar12} Narayanan, D., Krumholz, M.~R., Ostriker, E.~C., et al.\ 2012, \mnras, 421, 3127
\bibitem[Oman et al.(2021)]{oma21} Oman, K.~A., Bah{\'e}, Y.~M., Healy, J., et al.\ 2021, \mnras
\bibitem[Otter et al.(2022)]{ott22} Otter, J.~A., Rowlands, K., Alatalo, K., et al.\ 2022, \apj, 941, 93. 
\bibitem[Oxland et al.(2024)]{oxl24} Oxland, M., Parker, L.~C., de Carvalho, R.~R., et al.\ 2024, \mnras, 529, 4, 3651. 
\bibitem[Pan et al.(2024)]{pan24} Pan, H.-A., Lin, L., Ellison, S.~L., et al.\ 2024, \apj, 964, 120. 
\bibitem[Pessa et al.(2021)]{pes21} Pessa, I., Schinnerer, E., Belfiore, F., et al.\ 2021, \aap, 650, A134
\bibitem[Peng et al.(2015)]{pen15} Peng, Y., Maiolino, R., \& Cochrane, R.\ 2015, \nat, 521, 7551, 192.
\bibitem[Pettini \& Pagel(2004)]{pet04} Pettini, M., \& Pagel, B.~E.~J.\ 2004, \mnras, 348, L59
\bibitem[Piotrowska et al.(2020)]{pio20} Piotrowska, J.~M., Bluck, A.~F.~L., Maiolino, R., et al.\ 2020, \mnras, 492, L6
\bibitem[Piotrowska et al.(2022)]{pio22} Piotrowska, J.~M., Bluck, A.~F.~L., Maiolino, R., et al.\ 2022, \mnras, 512, 1, 1052. 
\bibitem[Querejeta et al.(2019)]{que19} Querejeta, M., Schinnerer, E., Schruba, A., et al.\ 2019, \aap, 625, A19. 
\bibitem[Rowlands et al.(2015)]{row15} Rowlands, K., Wild, V., Nesvadba, N., et al.\ 2015, \mnras, 448, 258
\bibitem[Rowlands et al.(2018)]{row18} Rowlands, K., Heckman, T., Wild, V., et al.\ 2018, \mnras, 480, 2, 2544. 
\bibitem[Saintonge et al.(2016)]{sai16} Saintonge, A., Catinella, B., Cortese, L., et al.\ 2016, \mnras, 462, 1749
\bibitem[Salim et al.(2007)]{sal07} Salim, S., Rich, R.~M., Charlot, S., et al.\ 2007, \apjs, 173, 267 
\bibitem[Salim et al.(2014)]{sal14} Salim, S., Lee, J.~C., Ly, C., et al.\ 2014, \apj, 797, 126
\bibitem[S{\'a}nchez et al.(2013)]{san13} S{\'a}nchez, S.~F., Rosales-Ortega, F.~F., Jungwiert, B., et al.\ 2013, \aap, 554, A58 
\bibitem[S{\'a}nchez et al.(2014)]{san14} S{\'a}nchez, S.~F., Rosales-Ortega, F.~F., Iglesias-P{\'a}ramo, J., et al.\ 2014, \aap, 563, A49
\bibitem[Tanaka et al.(2024)]{tan24} Tanaka, T.~S., Shimasaku, K., Tacchella, S., et al.\ 2024, \pasj, 76, 1. 
\bibitem[S{\'a}nchez et al.(2016a)]{san16a} S{\'a}nchez, S. F., P{\'e}rez, E., S{\'a}nchez-Bl{\'a}zquez, P., et al. 2016, \rmxaa, 52, 21
\bibitem[S{\'a}nchez et al.(2016b)]{san16b} S{\'a}nchez, S. F., P{\'e}rez, E., S{\'a}nchez-Bl{\'a}zquez, P., et al. 2016b, \rmxaa, 52, 171
\bibitem[S{\'a}nchez et al.(2018)]{san18} S{\'a}nchez, S.~F., Avila-Reese, V., Hernandez-Toledo, H., et al.\ 2018, \rmxaa, 54, 217
\bibitem[S{\'a}nchez(2020)]{san20} S{\'a}nchez, S.~F.\ 2020, \araa, 58, 99
\bibitem[S{\'a}nchez et al.(2021)]{san21} S{\'a}nchez, S.~F., Barrera-Ballesteros, J.~K., Colombo, D., et al.\ 2021, \mnras, 503, 1615
\bibitem[Sandstrom et al.(2013)]{san13} Sandstrom, K.~M., Leroy, A.~K., Walter, F., et al.\ 2013, \apj, 777, 1, 5. 
\bibitem[Sargent et al.(2014)]{sar14} Sargent, M.~T., Daddi, E., B{\'e}thermin, M., et al.\ 2014, \apj, 793, 19 
\bibitem[Sarzi et al.(2010)]{sar10} Sarzi, M., Shields, J.~C., Schawinski, K., et al.\ 2010, \mnras, 402, 2187 
\bibitem[Schawinski et al.(2014)]{sch14} Schawinski, K., Urry, C.~M., Simmons, B.~D., et al.\ 2014, \mnras, 440, 889 
\bibitem[Schmidt(1959)]{sch59} Schmidt, M.\ 1959, \apj, 129, 243 
\bibitem[Singh et al.(2013)]{sin13} Singh, R., van de Ven, G., Jahnke, K., et al.\ 2013, \aap, 558, A43 
\bibitem[Smethurst et al.(2015)]{sme15} Smethurst, R.~J., Lintott, C.~J., Simmons, B.~D., et al.\ 2015, \mnras, 450, 435 
\bibitem[Smirnova-Pinchukova et al.(2022)]{smi22} Smirnova-Pinchukova, I., Husemann, B., Davis, T.~A., et al.\ 2022, \aap, 659, A125
\bibitem[Stasi{\'n}ska et al.(2008)]{sta08} Stasi{\'n}ska, G., Vale Asari, N., Cid Fernandes, R., et al.\ 2008, \mnras, 391, L29 
\bibitem[Sun et al.(2020)]{sun20} Sun, J., Leroy, A.~K., Ostriker, E.~C., et al.\ 2020, \apj, 892, 148
\bibitem[Tacchella et al.(2022)]{tac22} Tacchella, S., Conroy, C., Faber, S.~M., et al.\ 2022, \apj, 926, 2, 134 
\bibitem[Teng et al.(2022)]{ten22} Teng, Y.-H., Sandstrom, K.~M., Sun, J., et al.\ 2022, \apj, 925, 1, 72.
\bibitem[Teng et al.(2023)]{ten23} Teng, Y.-H., Sandstrom, K.~M., Sun, J., et al.\ 2023, \apj, 950, 119
\bibitem[Thorp et al.(2022)]{tho22} Thorp, M.~D., Ellison, S.~L., Pan, H.-A., et al.\ 2022, \mnras, 516, 1462. 
\bibitem[Trussler et al.(2020)]{tru20} Trussler, J., Maiolino, R., Maraston, C., et al.\ 2020, \mnras, 491, 4, 5406.
\bibitem[Vazdekis et al.(2016)]{vaz16} Vazdekis, A., Koleva, M., Ricciardelli, E., et al.\ 2016, \mnras, 463, 3409.
\bibitem[Venturi et al.(2021)]{ven21} Venturi, G., Cresci, G., Marconi, A., et al.\ 2021, \aap, 648, A17. 
\bibitem[Villanueva et al.(2022)]{vil22} Villanueva, V., Bolatto, A.~D., Vogel, S., et al.\ 2022, \apj, 940, 2, 176. 
\bibitem[Villanueva et al.(2024)]{vil24} Villanueva, V., Bolatto, A.~D., Vogel, S.~N., et al.\ 2024, \apj, 962, 88. 
\bibitem[Visser-Zadvornyi et al.(2025)]{vis25} Visser-Zadvornyi, A.~I., Carstairs, M.~E., Oman, K.~A., et al.\ 2025, \mnras
\bibitem[Walters et al.(2022)]{wal22} Walters, D., Woo, J., \& Ellison, S.~L.\ 2022, \mnras, 511, 4, 6126. 
\bibitem[Wang et al.(2022)]{wan22} Wang, W.-H., Foucaud, S., Hsieh, B.-C., et al.\ 2022, \apjs, 260, 2, 54. 
\bibitem[Wan et al.(2024)]{wan24} Wan, J.~T., Tacchella, S., Johnson, B.~D., et al.\ 2024, \mnras, 532, 4002. 
\bibitem[Weinberger et al.(2017)]{wei17} Weinberger, R., Springel, V., Hernquist, L., et al.\ 2017, \mnras, 465, 3, 3291. 
\bibitem[Westfall et al.(2019)]{wes19} Westfall, K.~B., Cappellari, M., Bershady, M.~A., et al.\ 2019, \aj, 158, 6, 231. 
\bibitem[Wetzel et al.(2013)]{wet13} Wetzel, A.~R., Tinker, J.~L., Conroy, C., et al.\ 2013, \mnras
\bibitem[Wright et al.(2019)]{wri19} Wright, R.~J., Lagos, C. del P., Davies, L.~J.~M., et al.\ 2019, \mnras, Quenching time-scales of galaxies in the EAGLE simulations, 487, 3, 3740.
\bibitem[Wu et al.(2005)]{wu05} Wu, J., Evans, N.~J., Gao, Y., et al.\ 2005, \apjl, 635, L173
\bibitem[Wu(2021)]{wu21} Wu, P.-F.\ 2021, \apj, 913, 44.
\bibitem[Wyder et al.(2007)]{wyd07} Wyder, T.~K., Martin, D.~C., Schiminovich, D., et al.\ 2007, \apjs, 173, 293. 
\bibitem[Wuyts et al.(2013)]{wuy13} Wuyts, S., F{\"o}rster Schreiber, N.~M., Nelson, E.~J., et al.\ 2013, \apj, 779, 135
\bibitem[Yan \& Blanton(2012)]{yan12} Yan, R., \& Blanton, M.~R.\ 2012, \apj, 747, 61 
\bibitem[Zhang et al.(2023)]{zha23} Zhang, J., Li, Y., Leja, J., et al.\ 2023, \apj, 952, 6. 













\end{thebibliography}
\end{document}